\documentclass{jfm}
\usepackage{graphicx}
\usepackage{epstopdf, epsfig}
\usepackage{xcolor}
\usepackage{hyperref}

\shorttitle{Atomisation of a single levitated droplet in an acoustic trap}
\shortauthor{Saroj and Thaokar}

\title{Atomisation of an acoustically levitated droplet: Experimental observations of a myriad of complex phenomenon}

  \author{S. K. Saroj\aff{1}, R.M. Thaokar\aff{1}
  \corresp{\email{rochish@che.iitb.ac.in}},}

\affiliation{\aff{1}Department of Chemical Engineering, IIT Bombay, Mumbai
Maharashtra, 400076 India
}

\begin{document}

\maketitle

\begin{abstract}
We report the dynamics of a droplet levitated in a single-axis acoustic levitator.  The deformation and atomization behavior of the droplet in the acoustic field exhibits a myriad of complex phenomena, in sequences of steps. These include the primary breakup of the droplet through stable levitation, deformation, sheet formation, and equatorial atomization, followed by secondary breakup which could be umbrella breakup, bag breakup, bubble breakup or multistage breakup depending on the initial size of the droplet. The visualization of the interfacial instabilities on the surface of the liquid sheet using both side and top-view imaging is presented. An approximate size distribution of the droplet after a complete breakup is also provided. Lastly, an aggregation of the atomized smaller droplets is observed after the complete atomization. The primary breakup of the droplet precedes with a stable levitation of the droplet, when the acoustic force balances the downward gravity force and the resulting ellipsoidal shape of the droplet is a consequence of the balance of deforming acoustic force and the restoring surface tension force. The acoustic force changes with the change in the shape of the droplet leading to further deformation of the droplet, ultimately resulting in a highly flattened droplet, with a thin liquid sheet at the edge (equatorial region). The thinning of the sheet is caused by the differential acceleration induced by the greater pressure difference between the poles and the equator. As the sheet thickness reduces to the order of a few microns,  Faraday waves develop at the thinnest region (preceding to rim) which causes the generation of tiny-sized droplets perpendicular to the sheet. The corresponding, hole formation results in a perforated sheet that causes the detachment of the annular rim which breaks due to Rayleigh Plateau (RP) instability. The radial ligaments generated in the sheet possibly due to Rayleigh Taylor (RT) instability break into droplets of different sizes. The secondary breakup exhibits  We dependency and includes umbrella, bag, bubble, or multistage, ultimately resulting in complete atomization of the droplets. Both the primary and the secondary breakup of the droplet admit interfacial instabilities such as Faraday instability, Kelvin Helmholtz (KH) instability, RT instability, and RP instability and are well described with visual evidence. 
\end{abstract}

\section{Introduction}

A liquid droplet can be levitated in the air using acoustic radiation force generated by sound waves. The phenomenon, termed acoustic levitation, is one of the best methods for contactless transportation, deformation, and breakup of liquid droplets. The levitation of a droplet by sound waves finds several important applications in the field of material science (\cite{andrade2020acoustic}), biology (\cite{akkoyun2021potential}), and analytical chemistry (\cite{bayazitoglu1995experiments}). For example,   \cite{andrade2020contactless} developed a device based on acoustic fields for picking up Surface Mount Device (SMD) elements from one position and transporting and releasing them at another position without any contact with the solid object. \\

A liquid droplet can also undergo fragmentation due to a variety of reasons, such as the gravity fall of naturally occurring raindrops (\cite{villermaux2009single}),  due to the effect of airflow   (\cite{kirar2022experimental}) or due to impaction with solid surfaces (\cite{lohse2020double,villermaux2011drop}) etc. The atomization of the droplet has important applications such as internal combustion engines (\cite{lefebvre2017atomization}), aircraft engines (\cite{ju2017experimental}), cooling (\cite{breitenbach2018drop}), agricultural spray (\cite{law2001agricultural}), inkjet printing (\cite{el2017inkjet}), pharmaceutical and medical  devices (\cite{naidu2022novel}) etc.\\

When a droplet is placed in the acoustic waves, under specialized conditions, it could exhibit both levitation as well as fragmentation. The deformation dynamics of such a levitated droplet, which can eventually lead to breakup and fragmentation, depends upon the relative interplay of surface tension of the liquid-air interface and the acoustic pressure.  A levitated droplet is known to exhibit a variety of dynamical shape deformations depending on its size and ambient conditions. Such a  droplet is almost spherical at low sound intensities and becomes gradually flattens as the sound intensity is increased. An oblate spheroidal shape is formed at moderate sound intensities (\cite{andrade2019numerical}). \cite{hong2017dynamics} designed an acoustic levitation technique by generating an acoustic vortex field, which levitates a wide range of liquid or solid objects. This methodology has been suggested to control the deformation of water droplets. \cite{geng2014vertical} reported experimental and theoretical investigations of the vertical vibration of a droplet levitated in an acoustic field. The vibration of the droplet is harmonic due to a combination of both acoustic radiation and gravity forces. For an equal volume of the droplet, the vibration frequency varies with the change in shape of the levitated droplet. A good agreement between the theory and experiment for a droplet smaller than 2.5 mm was observed. \\

The acoustically levitated and stretched droplet can eventually undergo breakup when the inertial and pressure forces overcome the capillary forces.  \cite{aoki2020acoustically} experimentally investigated the dynamics of acoustically levitated droplets of water and ethanol.  Both stable and unstable conditions of the levitated droplet were predicted, through dependence on the diameter of the droplet and the sound pressure.
They suggested the following relation for the acoustic pressure $P_{max}$ beyond which stable levitation is not possible and a droplet can undergo breakup,
\begin{equation}
 P_{max}=\sqrt{\sigma \rho_G c^2 \Big(\frac{3.2}{d}-\frac{1.3\pi}{\lambda}\Big)}
\end{equation}
where, $\sigma$, $\rho_G$, c, d and $\lambda$ are surface tension, density of air, speed of sound, diameter of droplet, and wavelength of sound wave, respectively. The typical values of the wavelength of sound in their work were $\lambda= c/$f$= 8.5$ mm for $c=340 m/s$ and $f = 40 kHz$. Thus, the susceptibility of a levitated droplet to atomization increases as the liquid's surface tension decreases and the intensity of sound pressure increases. \cite{andrade2019numerical} in their numerical study, observed that a levitated water droplet disintegrates in an acoustic field when the separation distance between the emitter and reflector is less than resonance. They observed that the droplet deformed into a pancake shape when the separation distance between the emitter and reflector was higher than the resonance separation. More recently, the atomization of the acoustically levitated droplet was observed by some researchers (\cite{andrade2019numerical,aoki2020acoustically,naka2020breakup}). The atomization process includes flattening the droplet into a pan-cake shape, developing capillary waves, and splitting the droplet into a large number of smaller droplets (\cite{andrade2019numerical}),  \cite{aoki2020acoustically,naka2020breakup}. The rapid deformation of the droplet was suggested to be due to the Kelvin–Helmholtz instability, which creates interfacial deformation (\cite{naka2020breakup}). The measured velocity of the interface before atomization and the critical velocity from the Kelvin–Helmholtz instability were compared by \cite{naka2020breakup}. They reported an excellent match between experimental and theoretical velocity, whereby the atomization was attributed to the Kelvin–Helmholtz instability at the interfacial region of the droplet. \\

Acoustically levitated droplets have been used to investigate different types of phenomena. For example, 
\cite{kumar2010structural} studied the vaporization of a nano-silica encapsulated water droplet by heating with $CO_2$ laser in an acoustic levitator. They observed  bowl and ring shape of the droplet during evaporation in the acoustic field. Further, \cite{basu2012thermally} investigated the atomization of acoustically levitated fuel droplets by external heating. They reported that the onset of the KH instability at the droplet's edge induces droplet atomization. The small-scale atomization for diesel, kerosene, and bio-diesel was attributed to their higher surface temperature. Ethanol droplets were found to be more stable because of the smaller surface temperature due to their higher vapor pressure and latent heat of vaporization. In another study, it was observed that viscous droplets levitated in acoustic fields break into smaller droplets due to cavitation of a bubble inside the droplet (\cite{zeng2018jetting}), generated by localized heating inside the droplet using a laser source. The breakup is because of the radial acceleration of the droplet that arises due to the oscillation and expansion of the bubble inside the droplet. \\


The few above-discussed studies on the breakup of acoustically levitated droplets describe the deformation of the droplet into a thin liquid sheet before the onset of atomization. However, the events leading to the atomization of an acoustically levitated droplet are not very well investigated in the literature. On the other hand, the breakup of a liquid sheet has been extensively investigated for sheets produced by a liquid jet impacting on a solid circular surface (\cite{villermaux2002life,clanet2002life,bremond2007atomization}), two co-linear jets  \cite{mulmule2010instability}, oblique impingement of two liquid jets (\cite{dighe2019atomization}) and droplet impacting on solid surfaces (\cite{lohse2020double,josserand2016drop,villermaux2011drop}). The liquid jet impacting on a solid circular surface generates a circular liquid sheet of varying thickness in the radial direction. The liquid sheet eventually becomes unstable and disintegrates at the edge (\cite{villermaux2002life,clanet2002life}). The liquid sheet breakup mechanism is classified in two regimes depending on the Weber number $We_L= \frac{\rho_L U_{L}^2 d_j}{\sigma} $, where, $U_L$, $d_j$, and $\sigma$ are the jet velocity, jet diameter and surface tension of the liquid, respectively. The liquid sheet is considered to be in a smooth regime for $We_L < We_{Lc}$ and flapping regime for $We_L > We_{Lc}$, $We_{Lc} $ is the critical Weber number of the order of $10^3$. In the smooth regime, there is no surface instability and the liquid velocity of the sheet remains constant up to the sheet edge where a rim is formed. The sheet atomizes at the edge on account of the Rayleigh Plateau instability of the rim. Cusp formation could also be seen, wherein, the droplet ejection takes place when the centrifugal acceleration dominates over the capillary action (\cite{clanet2002life}) as the droplets move along the cusp region. In the flapping regime, the KH instability is known to develop because of the relative motion of the liquid wavefront in the air which causes a flag-like motion. This flag-like motion generates radial acceleration-deceleration at the edge of the sheet (rim), which triggers the Rayleigh Taylor instability and results in the disintegration of the edge (\cite{villermaux2002life}). The liquid sheet expansion and recoiling of a droplet impacting a surface happens at constant pressure. \cite{villermaux2011drop} argued that the generation of the ligaments is independent of the impacting velocity when the droplet impacts on the solid surface (\cite{villermaux2011drop}). The formation of ligaments from the sheet is attributed to a Rayleigh-Taylor instability, experienced by the decelerating rim, which itself tries to simultaneously undergo a Rayleigh-Plateau instability. \\

The effect of the acoustic field on the atomization characteristics of a liquid sheet generated by two co-linear jets was investigated by \cite{mulmule2010instability}. An acoustic speaker which was capable to deliver sound frequencies from 65 to 18000 Hz was located normal to the sheet. Surface ripples were seen to travel toward the edge of the sheet when the speaker was switched on. Hole formation in the sheet was also observed that resulted in the reduction in the sheet diameter. Both the dilatation and sinuous modes of the KH instability are known to couple at the lowest order under acoustic forcing and the growth rate of the dilatation mode is no longer negligible. In another similar study, the variation of liquid sheet thickness was found to play an important role in different instability characteristics. It was observed that while the aerodynamic effect dominates for smaller acoustic frequencies the thinning effect dominates at a higher frequency at a fixed Weber number (\cite{dighe2019atomization}). \cite{fang2022linear}  numerically investigated the instability of a liquid sheet under the influence of the acoustic field. The acoustic field was applied from both sides and perpendicular to the liquid sheet. They predicted that the unstable region is affected by the surface tension, viscosity,  and density of the liquid and air as well as the amplitude of the acoustic field. The instability was suggested to be due to the combined effect of KH waves and parametric oscillation. The sub-harmonic oscillations were shown to change to harmonic oscillations by an increase in the We due to the suppression of parametric oscillation while the parametric instability was found to overcome the KH instability at higher acoustic acceleration. The instability exhibited sensitive dependence on the thickness variation (\cite{jia2022experimental}).  \\

A liquid droplet can atomize by a mode of breakup, termed the bag breakup mode in literature, and is observed in several situations, such as a droplet falling freely in the air (\cite{villermaux2009single}), a droplet exposed to continuous horizontal jet (\cite{kulkarni2014bag}) and droplet exposed in swirling airflow (\cite{kirar2022experimental}) without any equatorial atomization. A freely falling droplet, for example, undergoes flattening followed by bag formation because of a radial pressure gradient along the droplet surface.  The temporal development of the radius of the droplet and the bag is exponential (\cite{kulkarni2014bag}). \cite{kirar2022experimental} observed “retracting bag breakup” for a droplet subjected to a swirling airflow. The interfacial instability of the drop was suggested to be because of the RT mechanism. \\

While there have been several studies on acoustic droplet levitation, deformation, and atomization, this work presents a comprehensive experimental investigation into a variety of aspects of the fate of a liquid droplet levitated at the node of a standing wave in an acoustic trap. Specifically, the work addresses levitation, droplet stretching, droplet thinning, primary atomization and various modes of secondary atomization of a liquid droplet levitated in an acoustic trap. The physics of these phenomena is discussed deriving from literature studies and our own studies, and several instabilities including the Kelvin Helmoholtz, Faraday, Rayleigh Plateau and Rayleigh Taylor instabilities, are observed and identified in experiments, discussed and comparisons made with respective theories. The work demonstrates a myriad of complex, nonlinear, coexisting phenomena that renders a single droplet into countless atomized droplets.

\section{Experimental details}

\subsection{Experimental setup}
\begin{figure*}
\centering
\includegraphics[width=1\textwidth]{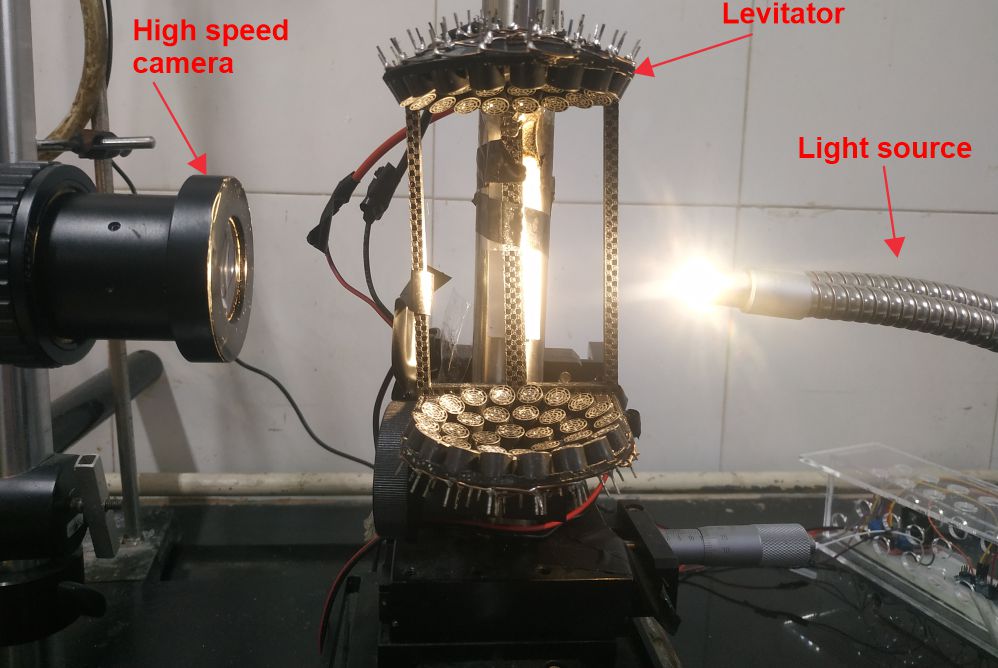}
\caption{\label{fig:setup} Experimental setup.}
\end{figure*}

A picture of the experimental arrangement is shown in Figure \ref{fig:setup}. The experimental setup included an acoustic levitation system, a fiber optics light source (Nikon, model: C-FI230), a high-speed camera (Vision Research, Phantom v12), and a computer. A light source was used for the illumination of the droplet and a high-speed camera was used to capture the images at a fixed time interval. The captured images were recorded on a computer. To capture top-view images another high speed camera Phantom VEO 710 camera was used. This camera had to be placed at 60 degrees to the vertical to enable proper visualization.

The acoustic levitator was constructed in accordance with the guidelines given by \cite{marzo2017tinylev}. It was made using two spherical cups of 4.5 cm, which are vertically placed at a center-to-center distance of 9 cm. The acoustic levitator consisted of 72 transducers, Arduino nano and a Dual H-Bridge motor driver. Each spherical cap had 36 transducers. The main components of the levitator are the transducers (CUI devices, 2223-CUSP-TR80-15-2500-TH-ND) that transform the electrical input into acoustic waves. The operation frequency of the transducer was equal to 40kHz. The corresponding wavelength of acoustic wave at 25$^{\circ}C$ equals 8.5 mm.  Arduino Nano was used to generate a high-speed square wave signal. The droplet was injected into the levitator using a syringe. The images were recorded from 4000 to 140k fps depending on the requirement of the experiment. The images were recorded with the camera focused on a certain droplet region to know the type of interfacial instability. The initial diameter of droplet is varied from 1.1 mm to 2.4 mm. The captured images were analyzed using the Image J software. The maximum pressure (pressure at anti-node) was estimated as 2400 Pa. For more details, the reader is advised to see the appendix section \ref{sound pressure}. The maximum pressure calculated by the simulation agrees closely with the pressure given by \cite{marzo2017tinylev}.

\subsection{Materials and method}
 The fluid properties of the droplet strongly influence the dynamics of an acoustically levitated droplet. The liquid droplet considered in this study was a mixture of 90\% ethanol and 10\% water. Before the experiment, the mixture was stirred continuously in an ultrasonic bath for 30 minutes to ensure proper ethanol-in-water mixing.  The corresponding surface tension and density of this mixture are 22 $mN/m$ and 830 $kg/m3$. The droplet was injected into the levitator using a syringe. The droplet then undergoes deformation followed by atomisation on the order of a millisecond. Therefore, the imaging was done at a frame rate of 4000 fps to 140000 fps.
 
\section{Results}\label{experiments}
We present the experimental results and observations in this section, followed by a discussion on possible mechanisms for these observations in the next section. The droplet gets levitated only for critical acoustic parameters. Since the acoustic wavelength is $8.5 mm$, the upper limit of the droplet size that can levitate in this levitator is equal to 4.4 mm. We only consider such levitated droplets in this study. The experimental observations include stable and unstable levitated droplets depending upon the Weber number. The terms "stable droplet" or "unstable droplet" are used for a levitated droplet that either resists (stable) or undergoes (unstable) atomization, respectively. Our primary focus in this article is to understand the instabilities associated with the atomisation of an acoustically levitated unstable droplet. \\

\begin{figure*}
\centering
\includegraphics[width=1\textwidth]{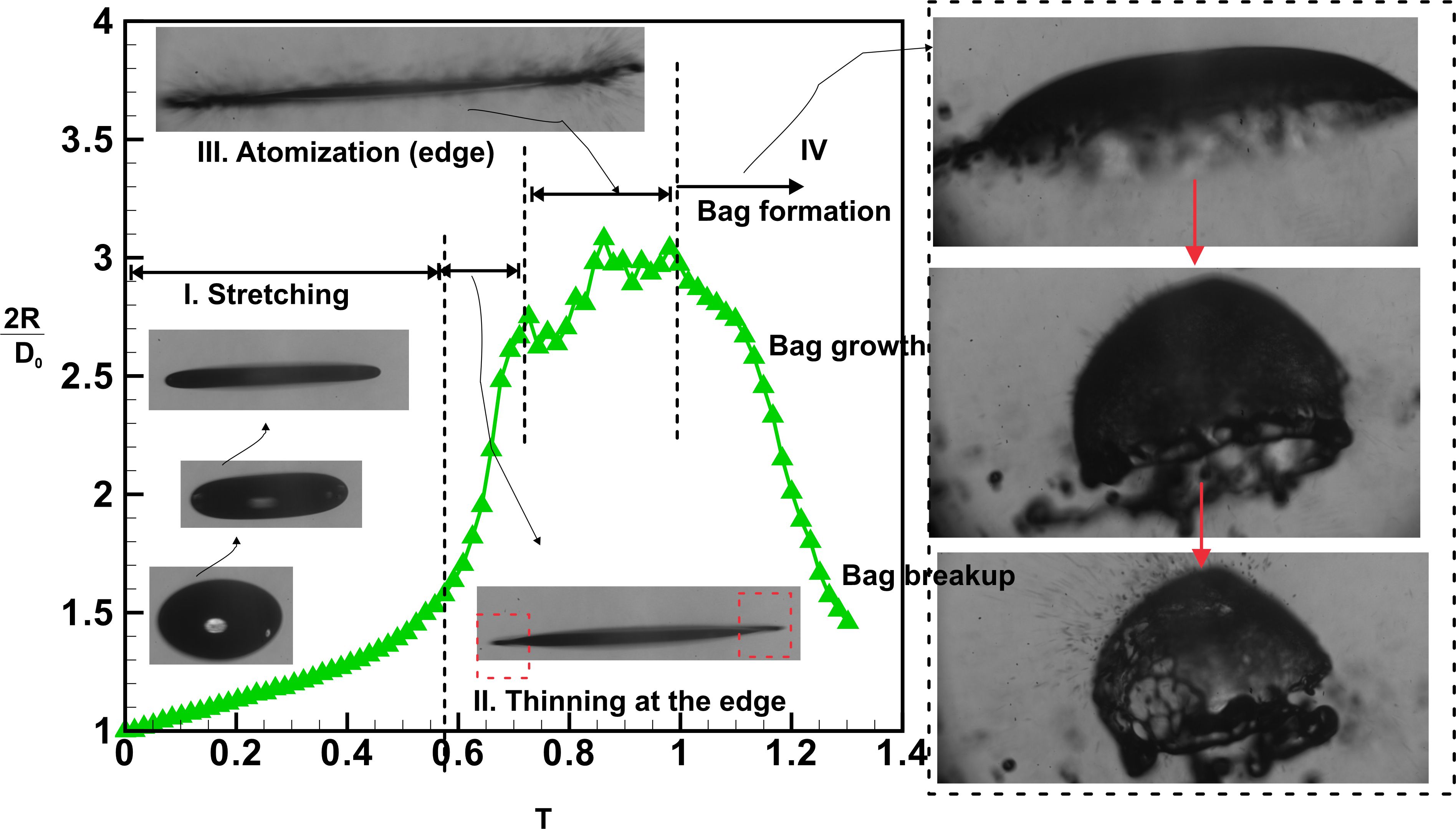}
\caption{\label{fig:stages} Different regimes of the droplet breakup. }
\end{figure*}

\begin{figure*}
\centering
\includegraphics[width=0.8\textwidth]{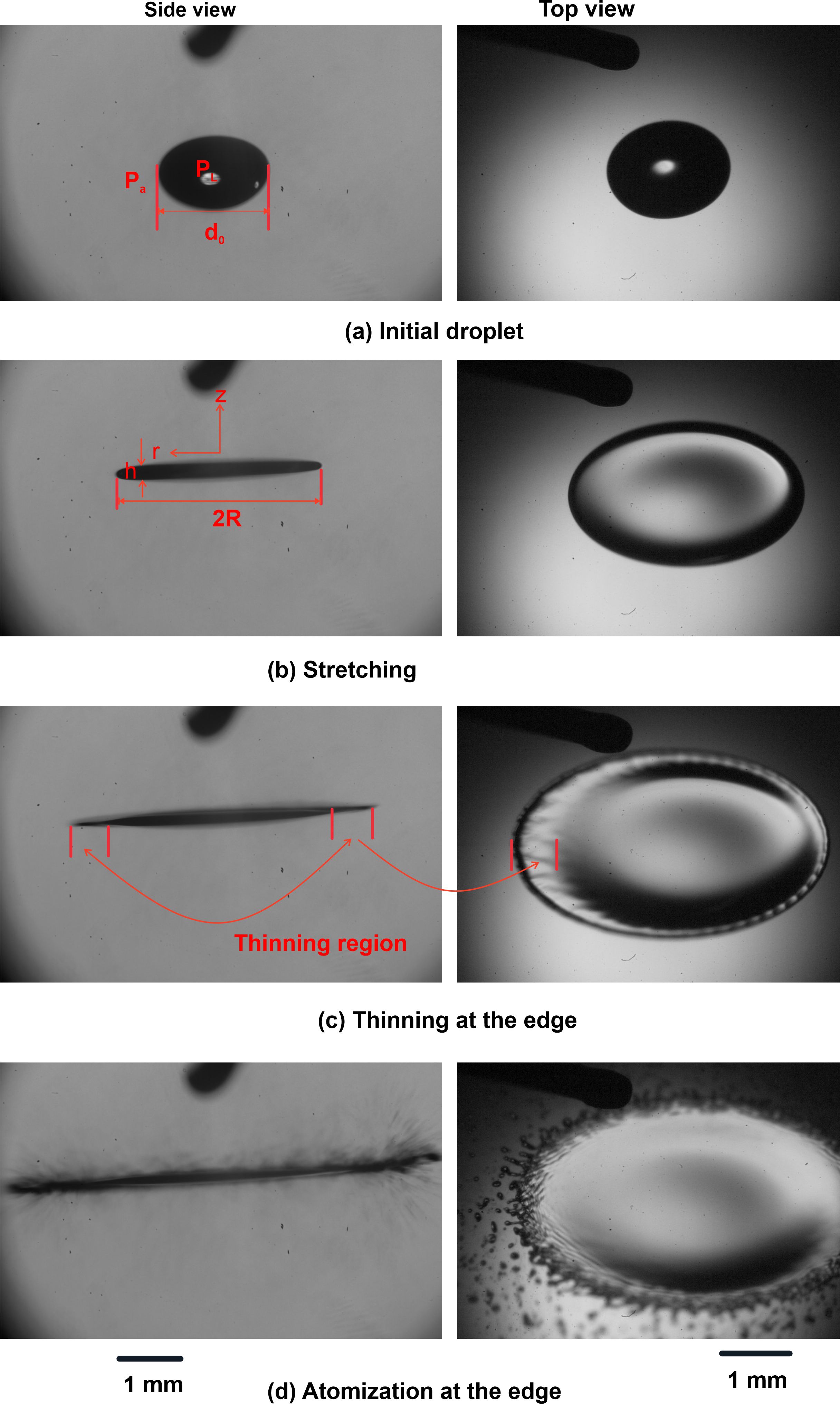}
\caption{\label{fig:regimes} Side view and top view images at different regimes of the droplet: (a) initial droplet, (b) droplet flattened into a flat disc shape, (c) thinning at the edge, and (d) atomisation at the edge). The side view images are captured at 6200 fps and the top view images were captured at 14485 fps. }
\end{figure*}

The radial expansion of a droplet in an acoustic field is classified into different regimes, as illustrated in Figure \ref{fig:stages}. For better clarity, the corresponding top and side view images are also shown in Figure \ref{fig:regimes}.  Droplet flattening occurs when an injected droplet is levitated at the node of a standing wave. The droplet first transforms from a spherical shape to a flat disc shape (see \ref{fig:regimes} (a) and (b)) and is termed the stretching regime. This is followed by the thinning regime that refers to the thinning of the edge region (equatorial region of the droplet) ((see \ref{fig:regimes} (c)). Atomization at the edge refers to the disintegration of the liquid sheet at the edge (see \ref{fig:regimes} (d)) Bag formation occurs at the end of edge atomization during the deceleration of the atomized sheet. The bag breakup occurs in 4 different ways: umbrella breakup, bag breakup, bubble breakup, and multistage breakup. The mechanism of atomization includes several interfacial instabilities in various breakup regimes. In the following sub-sections, we detail the observations made at different stages from the time a droplet is introduced in an acoustic trap to the point where it undergoes complete shattering and disintegration.

\subsection{Droplet Levitation in acoustic trap and critical We}

A droplet when injected at the node of a standing wave in an acoustic levitator does center-of-mass oscillations while simultaneously undergoing deformation, to finally reach the equilibrium position (node of the acoustic field). Experiments show that the oscillations decrease with an increase in the size of the droplet. Once at the node, the droplet remains steady and admits deformation at low We. In this case, the acoustic force is not sufficient to overcome the surface tension force and therefore the droplet does not undergo breakup. The oscillations of the droplet stop after a few seconds of injection as the acoustic and surface tension forces completely balance each other. The static equilibrium shape of the droplet depends on the Weber number (We) which is defined as $We = \frac{\rho_a U^2 d_0}{\sigma}$, where $\rho_a$, U, $d_0$, and $\sigma$ are the density of air, average acoustic velocity, initial diameter of the droplet and surface tension of the liquid, respectively. Note that the velocity $U=\frac{P_0}{\rho_0 v_0}  \frac{1}{\sqrt{2}}$ and is of the order of $4.09$ m/s since $P_0$ is of the order of 2400 Pa, $\rho_a$ and $v_0$ of the order 1.22 $kg/m^3$ and 340 $m/s$ respectively.

\begin{figure*}
\centering
\includegraphics[width=1\textwidth]{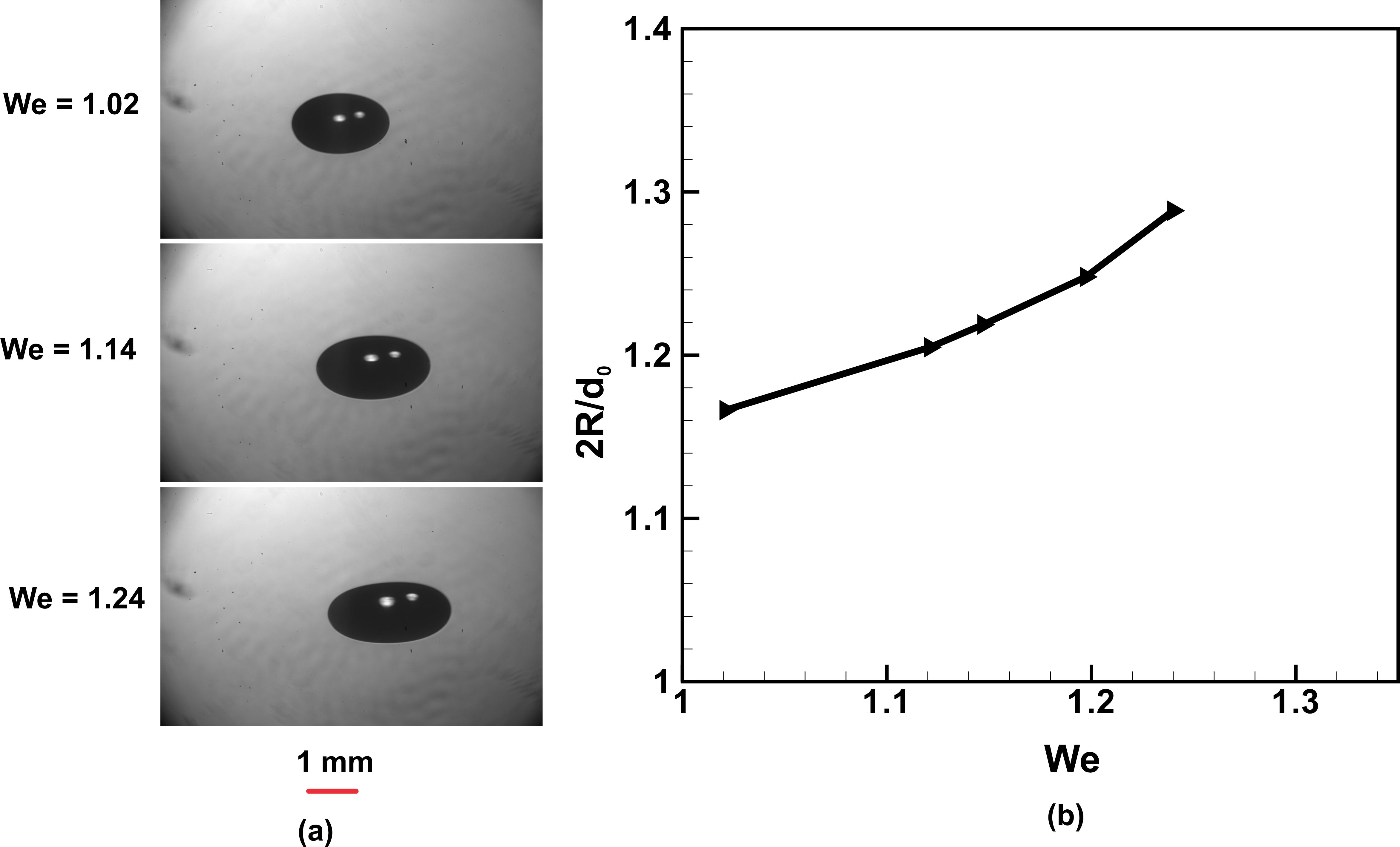}
\caption{\label{fig:stabledroplet} Static deformation of the droplet in the standing wave: (a) captured images at different We and (b) $\frac{2R}{d_0}$ vs We.}
\end{figure*}

Figure \ref{fig:stabledroplet} (a) shows the experimental data for the steady deformation of a stable levitated droplet (without breakup) placed in an acoustic standing wave. The droplet assumes an oblate spheroidal shape and the corresponding variation of its deformation is represented by ($\frac{2R}{d_0}$) at different We are presented in Figure \ref{fig:stabledroplet} (b). Here $2 R$ is the length of the principal-major axis of the oblate spheroidal droplet (R termed as radius of the oblate droplet in this work). The value of $2R$ and thereby the deformation of the droplet increases with increasing  We along with a corresponding reduction of the length of the principal minor axis (termed thickness $h$ here), such that  $\frac{2R}{d_0}$ $\simeq$ $We^{1/2}$ (figure \ref{fig:stabledroplet} (b)), indicating the prevalence of non-linearity. A similar observation has also been reported in numerical calculations (\cite{lee1991static}). There exists a critical We=1.36 (in our experiments) beyond which the droplet can no longer attain steady deformation (termed stable droplet), but instead continues to elongate (termed unstable droplet) ultimately leading to its breakup.

\subsection{Droplet flattening}
\begin{figure*}
\centering
\includegraphics[width=1\textwidth]{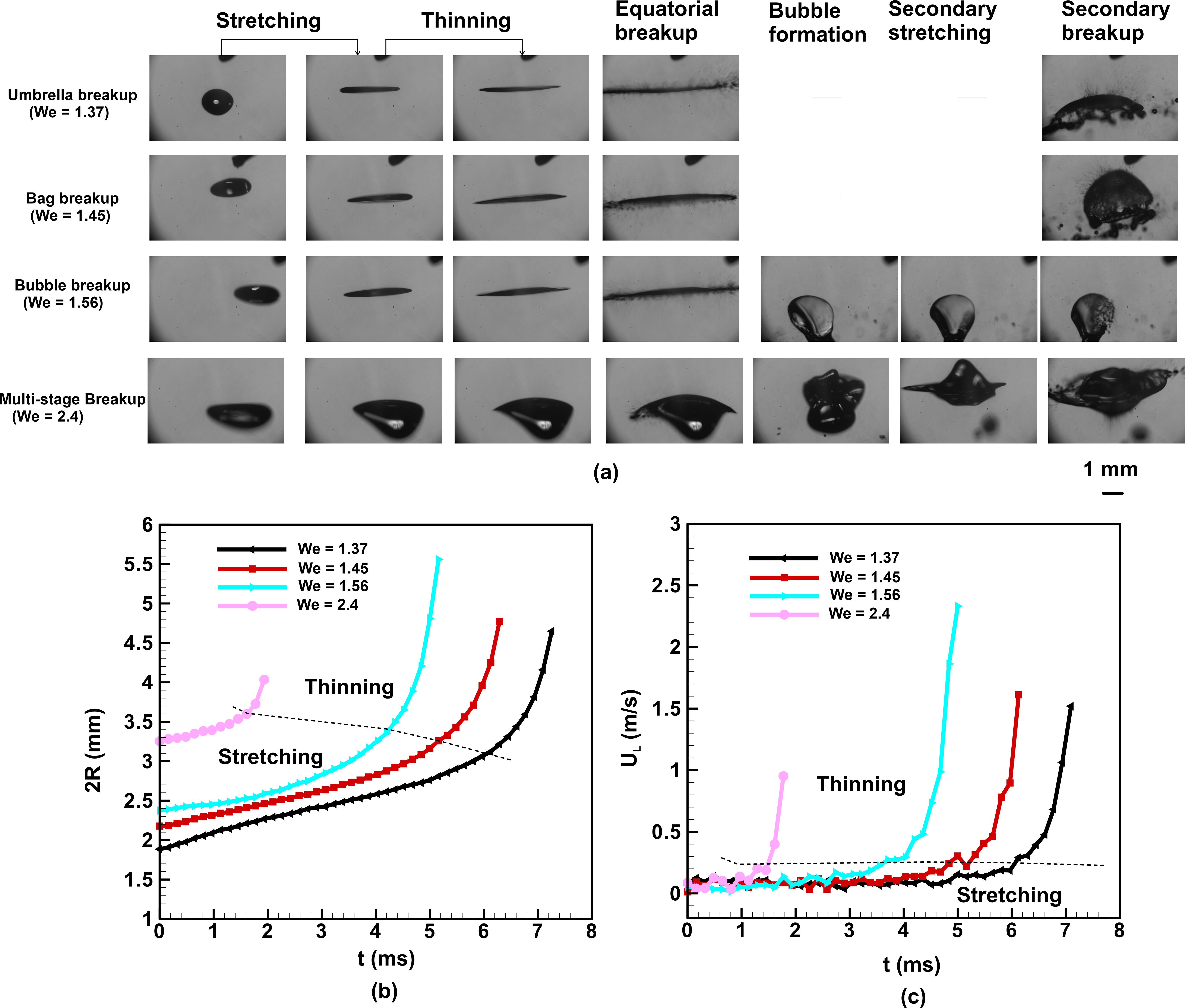}
\caption{\label{fig:stretching+thinning} (a) Side view visualization images at different stages, (b)  evolution of the diameter of the droplet and (c) corresponding rim velocity at different time for various We. Note: the images have been recorded at 6200 fps and the respective video are available at: (i)  {\color{blue}(\href{https://drive.google.com/file/d/1uNpBseCEywiqRO6D_WsCrp_e4ppnc0xR/view?usp=share_link} {umbrella breakup})}, (ii) {\color{blue} (\href{https://drive.google.com/file/d/1i1mrZJKZwyNsm6pOnpSG5HqDsv2HNhiF/view?usp=share_link} {Bag breakup})}, (iii)  {\color{blue}(\href{https://drive.google.com/file/d/1KhiVZ3t9Kfe6gqWDFHN6IMdWYKmOCfWR/view?usp=share_link} {Bubble breakup})} and (iv)   {\color{blue}(\href{https://drive.google.com/file/d/1IVYYClv2qbpRfHimt0DpvcbIcHI9mdKX/view?usp=share_link} {Multistage breakup})}.}
\end{figure*}

\begin{figure*}
\centering
\includegraphics[width=1\textwidth]{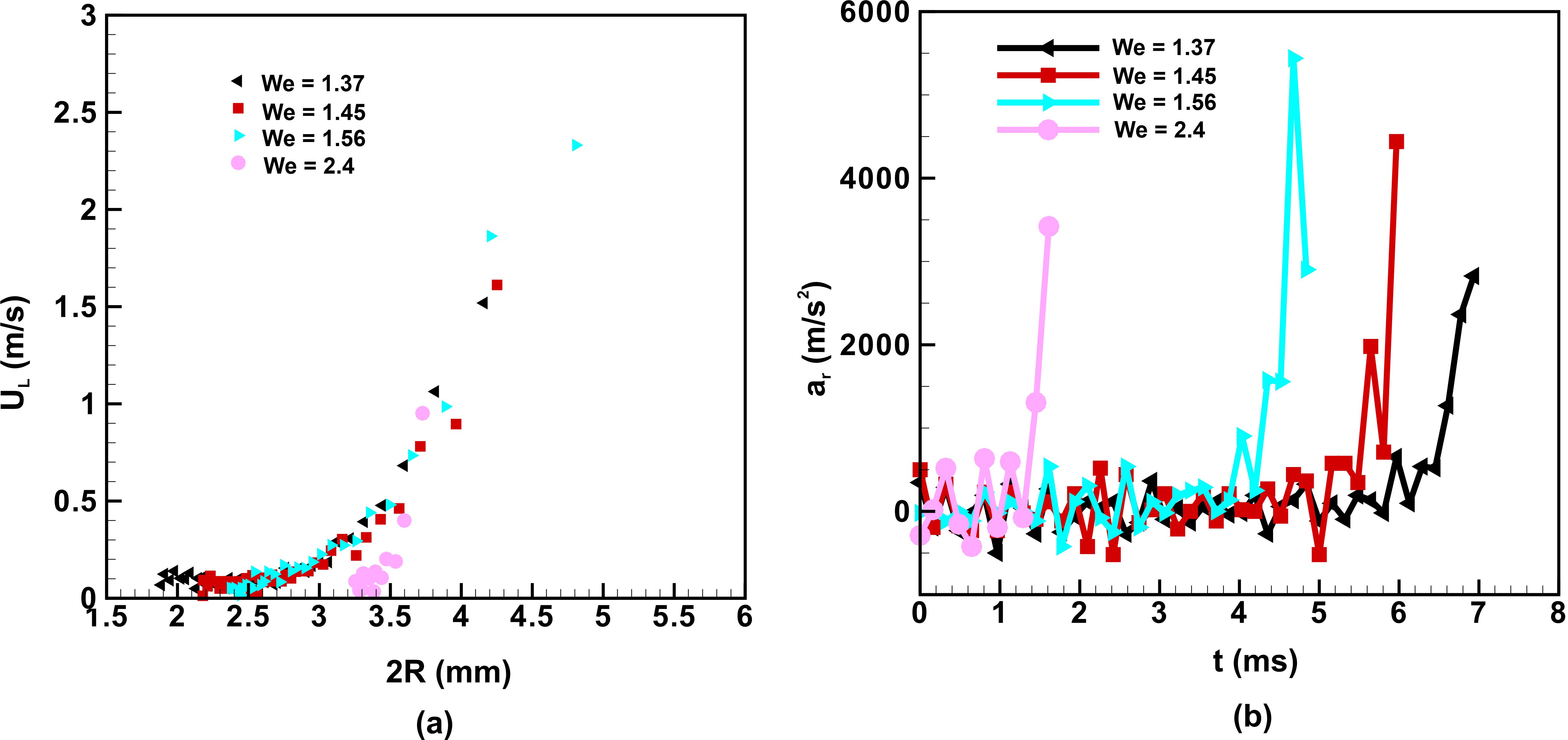}
\caption{\label{fig:ulvs2r} Rim propagation velocity ($U_L$) as a function of 2R at different We corresponding to figures \ref{fig:stretching+thinning} (b) and (c).}
\end{figure*}

The droplet, beyond the critical We$>$1.36,  changes from a spherical shape to an oblate shape, and its radius $R$ continues to increase with time as the droplet is squeezed at the poles.  The flattening of the droplet results in a highly flattened pan-cake like liquid sheet.  Figure \ref{fig:stretching+thinning} (a) shows the visualization images for different types of breakup modes depending on the We. The corresponding evolution in diameter of the droplet (2R) as a function of time at different We is presented in Figure \ref{fig:stretching+thinning} (b) which includes both the regimes, i.e. stretching as well as thinning (before the initiation of atomization). The stretching and thinning regimes are separated by a dotted line.  The variation of rim velocity ($U_L$), during radial expansion, as a function of 2R is shown in Figure \ref{fig:ulvs2r}. The range of rim velocity, admitted in the thinning regime, is approximately of the same order for all the We in the stretching regime. The onset of the thinning regime can be easily seen in the $U$ vs $t$ plot  (figure \ref{fig:ulvs2r}) and the thinning is initiated earlier (shorter times) at higher We. \\

A better understanding of the droplet stretching in the stretching regime is seen in figure \ref{fig:stretching+thinning} (b) and (c) which shows the evolution of diameter (2R) as a function of time for different We for the stretching (flattening) regime. The stretching time reduces with an increase in the We (initial diameter of the droplet). There is a good agreement between the experiment and deformation predicted by the equation \ref{phi}, the theory and mechanism of which are discussed in the next section. \\


Note that the droplet is placed at the node of a standing sound wave, where the pressure is zero and the associated velocity field in the air is maximum. This velocity field of air, associated with the sound wave, alternates sinusoidally in time and acts as the ambient air velocity field in which the droplet is placed. When the droplet is placed at the node, it can be likened to a problem of a droplet placed in an oscillatory uniform velocity field, wherein the droplet disturbs the ambient velocity field. The resulting oscillatory velocity field is such that when air goes around the droplet, it creates stagnation points at the poles, generating positive pressure fields. Importantly, at the equator, the air velocity is maximum, although oscillatory, which creates a non-oscillatory low-pressure region at the equator, consistent with the Bernoulli equation. It is this low pressure, created at the equator of the droplet, that leads to the flattening of the droplet. This is quantitatively demonstrated in the next section. 

 \subsection{Droplet Thinning}
 \label{section:droplet thinning}

As the droplet stretches in the flattening regime, the pressure difference between the pole and equator ($ P_p - P_E$) keeps increasing (details can be seen in appendix figure \ref{fig:comsol}), and then shows a nonlinear increase, which results in a sudden increase in the stretching rate ( with a corresponding increase in radial velocity), and thereby resulting in a sudden reduction in thickness at the edge of the liquid sheet. \cite{shi1995dynamics} made a similar observation using boundary element calculations, claiming that the inception of rim formation at the edge of the droplet, could lead to thinning of the region just preceding the rim. This is akin to jet thinning in Newtonian and non-Newtonian jets. For more clarity, we simulated the sound pressure distribution around the droplet of different aspect ratios. The simulation details are provided in Appendix \ref{acoustic thinning}. The aspect ratio is chosen in accordance with the experimental findings. It is clear that the compression pressure increases at the poles while it decreases at the equator during flattening (increase in $R/h$ ratio). The nonlinear increase in the equatorial velocity, due to the flattening of the droplet, creates a corresponding nonlinear decrease in the equatorial pressure, leading to the onset and sustenance of the thinning of the droplet.  The non-linearity is also reflected in the increase in the stretching factor $f$ with We, in the model to be discussed later. \\

\begin{figure*}
\centering
\includegraphics[width=1\textwidth]{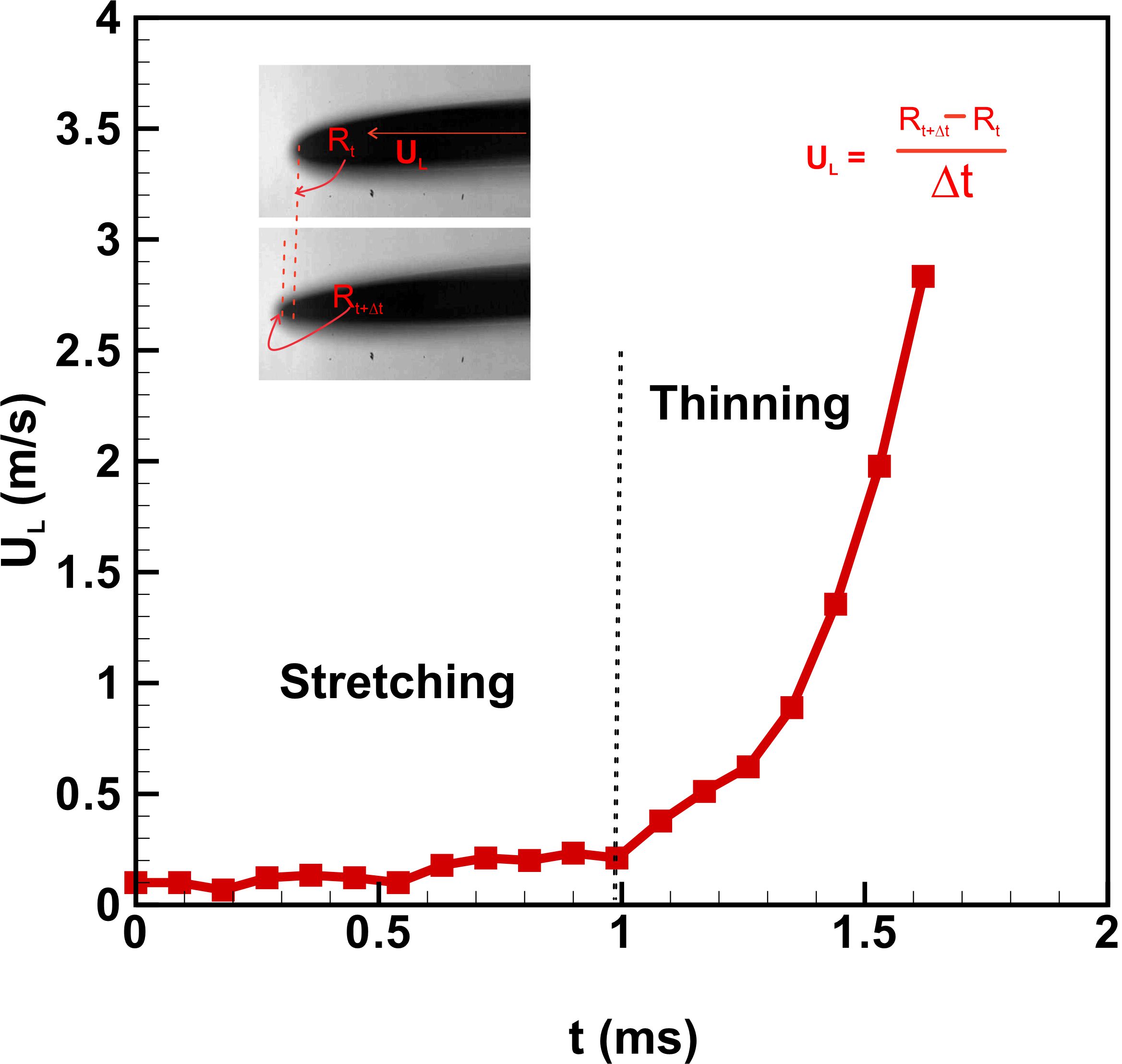}
\caption{\label{fig:vel} Velocity of the radially expanding droplet before initiation of breakup. Note: the images are recorded at 1100 fps with focusing on the equatorial region of the droplet. The velocity is calculated for the 1.5 ms before initiation of breakup which covers both stretching as well as thinning regime.}
\end{figure*}

 Figure \ref{fig:vel} shows the velocity of the liquid-air front at the equatorial region during radial expansion. The velocity is determined by extracting digital images recorded by a high-speed camera at 11000 frames per second for 1.82 ms before the atomization commences. The droplet is positioned in such a way that the edge region is completely captured during radial propagation of the liquid-air wavefront. The velocity is shown for both the stretching and thinning regimes. In the stretching regime, the velocity does not change significantly (it increases gradually), whereas it is faster in the thinning regime.\\

 Thus, the liquid sheet thins out near the equatorial region to replicate a membrane. The inertial force due to radial liquid velocity $U_L$ is stabilized by the surface tension force, which causes the accumulation of the liquid at the edge resulting in the formation of the rim as shown in Figure \ref{fig:rim} (a). The rate of change of momentum at the rim is resisted by the surface tension force (\cite{culick1960comments,villermaux2002life}). A balance of the inertial acceleration of the sheet and the interfacial tension, $\rho_L U_{LTC}^2 h = 2 \sigma$, leads to the Taylor Culick velocity $U_{LTC}= \sqrt{\frac{2 \sigma}{\rho h}}$.
According to the \cite{villermaux2002life}, a liquid sheet propagates in the air without initiation of breakup when $ \frac{\rho_L U_L^2 h}{\sigma}> 2$. For a constant velocity of stretching of a sheet, the sheet breaks when the thickness reduction is of the order of $2 \sigma/(\rho_L U_L^2)$. For example, in a liquid sheet formed by coaxial jets, the liquid velocity $U_L$ of the sheet remains constant, and a steady state is typically considered. The thickness $h$ decreases with $r$, the radial extent of the sheet, since mass conservation demands that $h=Q/(2 \pi r U_L)$. In this case, the liquid sheet breaks when the thickness h reaches that given by eqn \ref{TC}. 
 
\begin{equation}
      h = \frac{2 \sigma}{\rho_L U_L^2}
      \label{TC}
\end{equation}

In the present case of droplet breakup in an  acoustic field, the liquid velocity increases exponentially (see Figure \ref{fig:vel}) in the thinning regime. The inertial force $\rho_L U_L^2 h$ can therefore decrease to a value of $2 \sigma$, the Taylor Cullick criteria of breakup, only if there is a correspondingly faster decrease in the thickness of the sheet.

 At the edge, the rim thickness is greater than in the preceding regions due to the dominance of surface tension. As the rim and thereby the sheet accelerates due to lower pressure at the equator, $U_L$ increases with time, it draws an accelerating fluid from the region just behind the rim, in the equatorial region. The acceleration results in thinning since the local liquid is accelerated, and the mass balance then demands that the sheet reduces in thickness. This is a reinforcing process, where the high velocity in the thin region, further increases the radial extent of the thin region. While this may appear akin to the jet thinning mechanism it should be noted that the capillary pressure due to azimuthal curvature that plays an important role in the jet thinning mechanism is absent in the sheet thinning process. The mechanism involved in thinning of the liquid sheet at the equatorial region is qualitatively discussed in Appendix section \ref{acoustic thinning}. \\

\begin{figure*}
\centering
\includegraphics[width=1\textwidth]{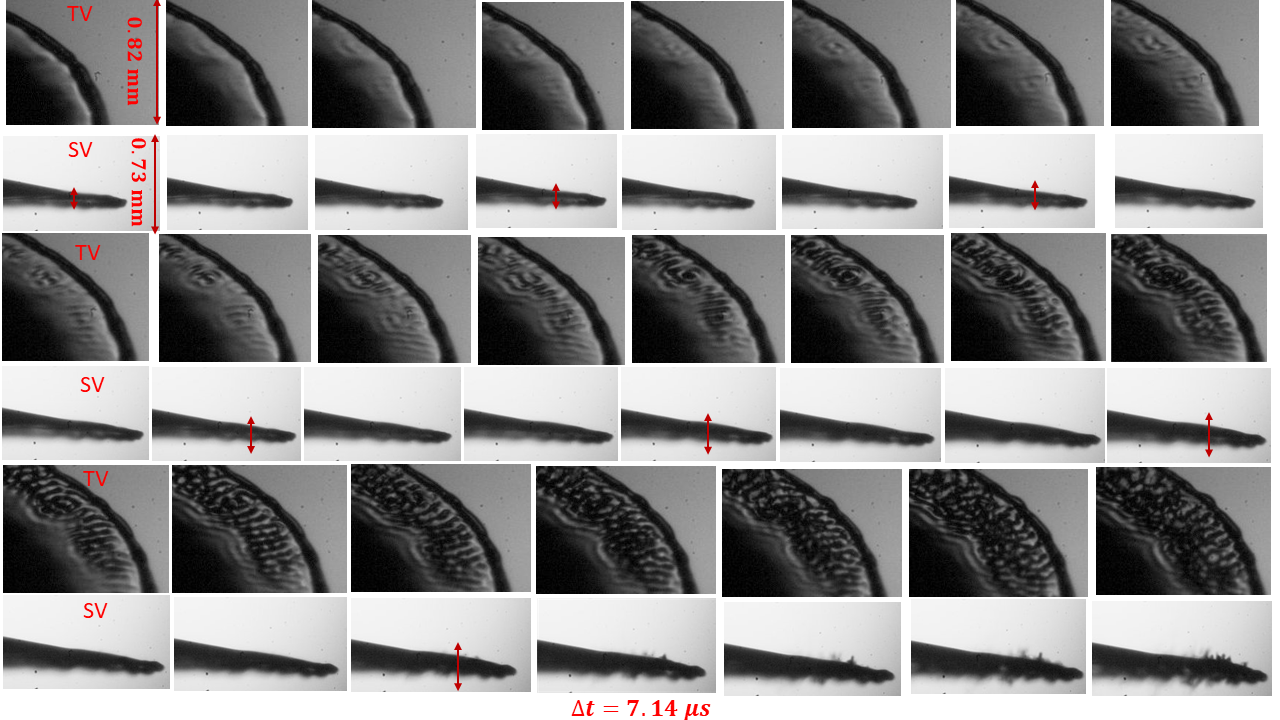}
\caption{\label{fig:1} Instantaneous top and side view digital images recorded at 140000 frames per second in the thinning regime. The images are recorded focusing at the equatorial region. Here, TV and SV indicate the top view and side view, respectively. The corresponding top and side view video can be found at: (a) {\color{blue}(\href{https://drive.google.com/file/d/1cu-lY1cbtVEn5Vc0Hfw-i-PYJOfh5u_j/view?usp=share_link}{video: top view}}) and (b) {\color{blue} (\href{https://drive.google.com/file/d/18YgXvJytOEKr_p3F5mxpWnGN7Y6x6Zav/view?usp=share_link}{video:side view})}.}
\end{figure*}

\begin{figure*}
\centering
\includegraphics[width=1\textwidth]{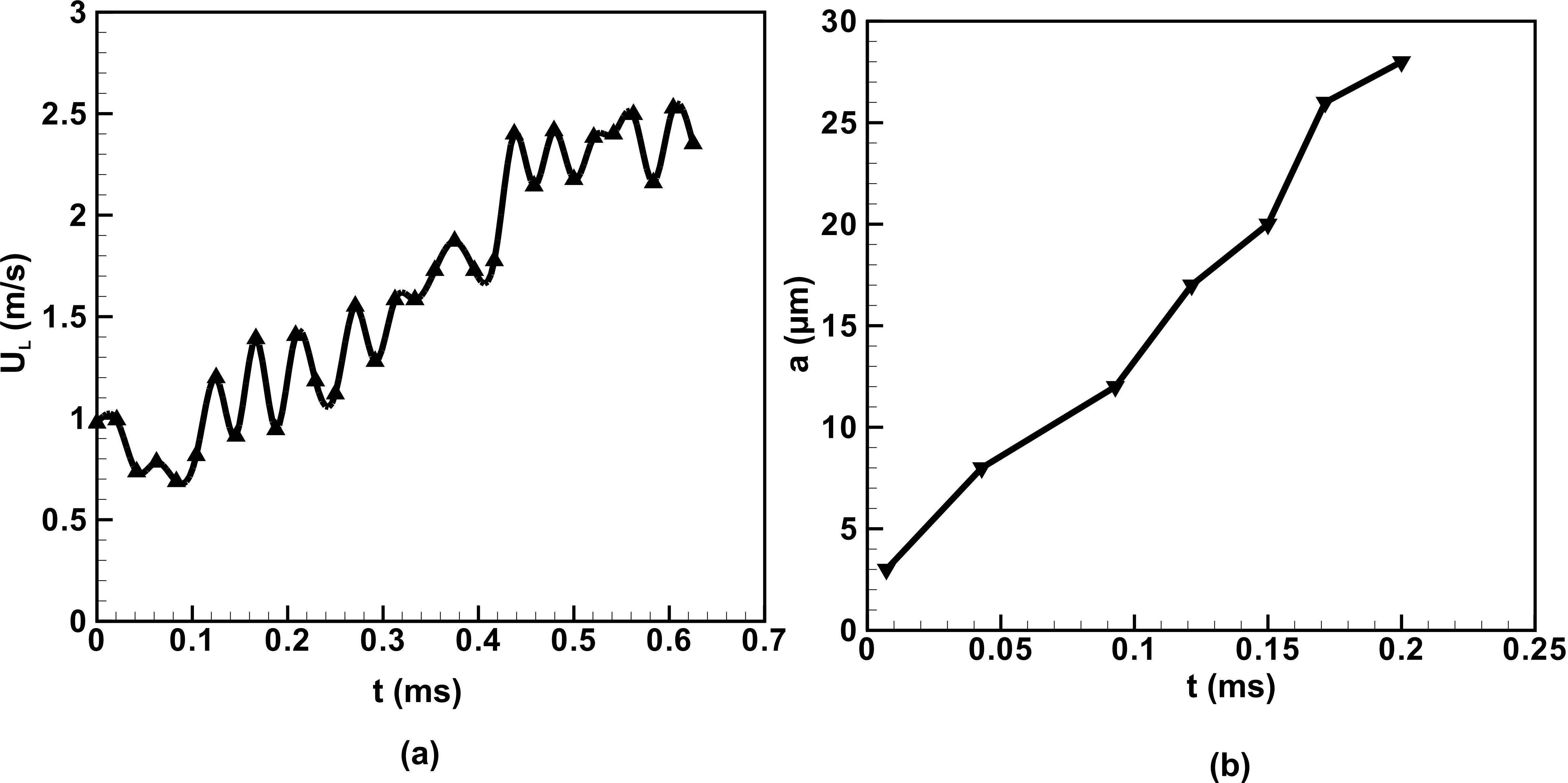}
\caption{\label{fig:velandampl} Rim velocity (a) and amplitude of vibrated liquid sheet (b) with time during thinning regime. Velocity is calculated from the side view results presented in figure \ref{fig:1}. Note: velocity is measured for 0.7 ms before the initiation of breakup and amplitude (vertical displacement of vibrating liquid sheet at thin region) is measured for 0.22 ms before the initiation of breakup.}
\end{figure*}

The velocity and acceleration in the thinning regime are actually periodic, as confirmed by experimental measurements obtained at 140k fps from top and bottom view images. The standing wave has an acoustic frequency of 40k Hz. Therefore, we captured the images at an fps higher than the frequency of the standing wave. Figure \ref{fig:1} shows the dynamics of the radial expansion of the liquid sheet in the thinning regime. Both the top and side view digital images at time intervals of 7.14 $\mu$ s are shown in Figure \ref{fig:1}. The rim velocity ($U_L$) and variation in the vertical amplitude with time in the thinning regime are shown in Figure \ref{fig:velandampl} (a) and Figure \ref{fig:velandampl} (b), respectively. The velocity $U_L$ in figure \ref{fig:velandampl} (a) when compared with the velocity shown in Figure \ref{fig:vel} indicates period averaged (40 kHz)  radial propagation in the latter, and therefore without any visible velocity fluctuations, especially in the thinning regime. This is essentially due to the lower frame speed in those experiments, 11000 fps, much smaller compared to the acoustic vibration frequency. The linear increment in the amplitude occurs in the unstable region (see Figure \ref{fig:velandampl} (b)), commensurate with the net acceleration of the drop edge. It should be noted here that $U_L$ in figure \ref{fig:velandampl} does show periodic deceleration of the drop edge.\\

{\color{red} The radial deformation time t (stretching+thinning) decreases with time as shown in table Table \ref{table:wevsteb} }

\subsection{Formation of Faraday waves at the thinned equatorial region before rim breakup}
\begin{figure*}
\centering
\includegraphics[width=1\textwidth]{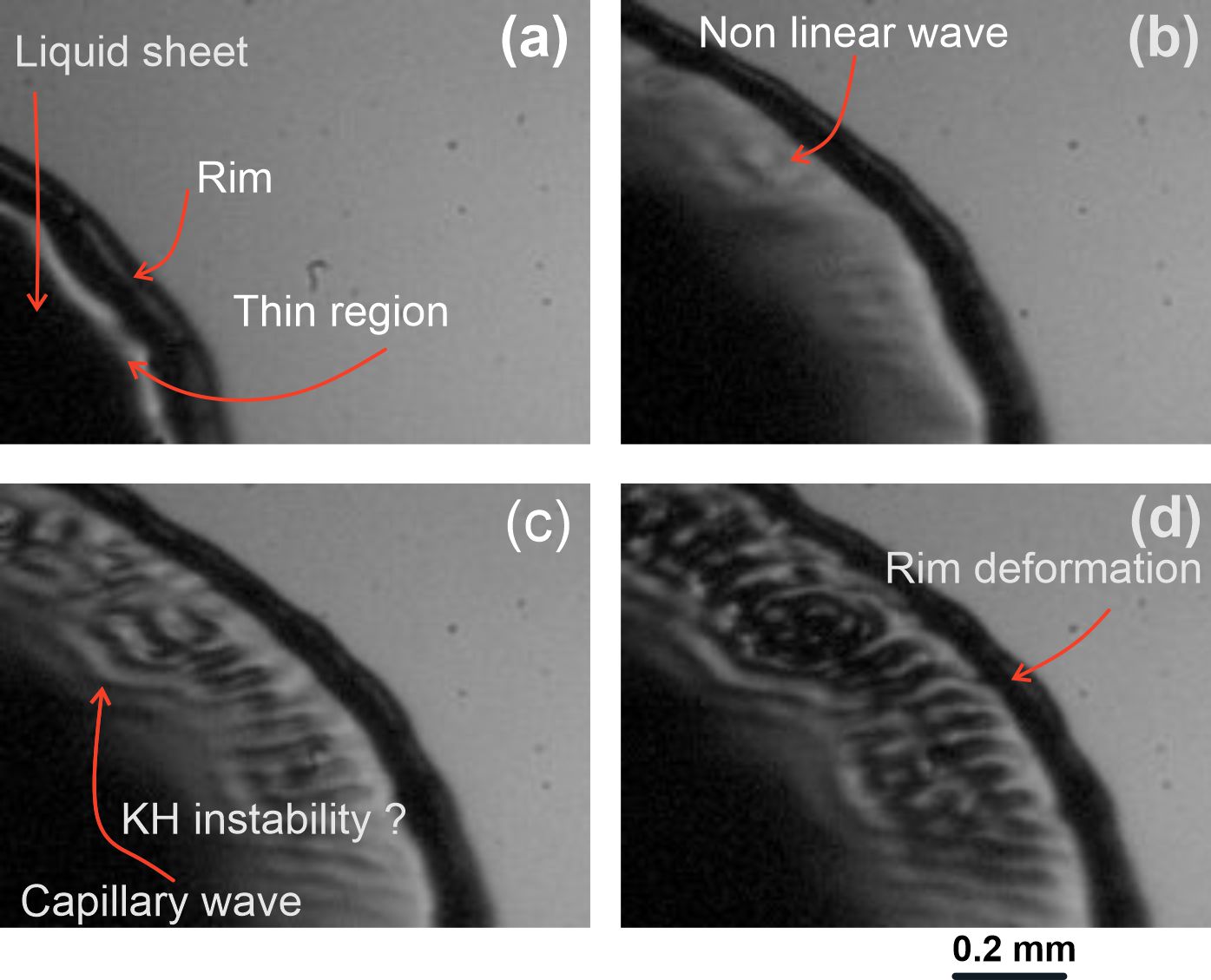}
\caption{\label{fig:rim} (a) rim formation, (b)  initiation of  nonlinear instability, (c) formation of capillary wave and (d) increasing of amplitude .}
\end{figure*}

The rim and thin region of the liquid sheet is indicated in Figure \ref{fig:rim} (a). A non-linear wave appears on the surface of the liquid sheet at the thinnest portion (preceding to the rim) (see Figure \ref{fig:rim} (b)). This non-linear pattern of the wave on the surface of the liquid is reported in the literature for wave generation due to the vertically oscillating liquid bath (\cite{oza2014pilot,tambasco2016onset,bush2015pilot}). The wave pattern that has a strong resemblance to the Faraday waves grows on the liquid sheet and induces a capillary wave (due to vertical oscillation of the liquid sheet) (see Figure \ref{fig:rim} (c)). The liquid sheet vibrates in the vertical direction which is shown in side view images of Figure \ref{fig:1}. 

\subsection{Breakup due to Faraday instability at the thinned equatorial region before rim breakup}
\label{Fragmentationedge}
\begin{figure*}
\centering
\includegraphics[width=1\textwidth]{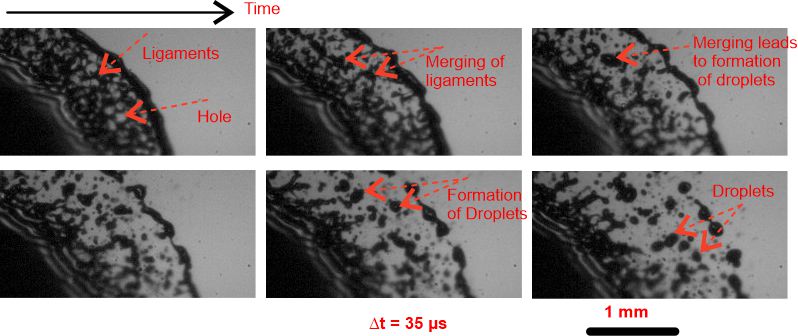}
\caption{\label{fig:merging} The merging of the interconnected ligaments during hole creation due to ejection of tiny sized droplets perpendicular to the liquid sheet at equatorial region .}
\end{figure*}

The vertical vibrations cause alternate radial acceleration and deceleration of the rim resulting in azimuthal deformation of the rim shown in Figure \ref{fig:rim} (d). 
The radial propagation of the liquid sheet with increasing velocity leads to a continuous reduction in the thickness. These azimuthal deformations propagate backward into the sheet, causing azimuthal perturbations in the Faraday waves in the thin region of the liquid sheet. Our observation suggests that a liquid sheet of a very small thickness, subject to an acoustic field shows much enhanced vertical vibrations. The amplitude of the vibration continues to increase linearly with time along with a reduction in the thickness (see Figure \ref{fig:velandampl} (b)). The Faraday waves now develop into a Faraday instability on the liquid sheet's surface, and a strong ejection of very tiny droplets is seen perpendicular to the surface of the liquid sheet. The ejection of these droplets from the liquid surface produces holes in the liquid sheet (see Figure \ref{fig:breakup} (a)). The perforation of the liquid sheet causes the ligaments to merge and break into droplets (Figure \ref{fig:breakup} (b)) thereby weakening the attachment of the rim with the flattened droplet. The pictorial representation of the ligament merging during perforation of the liquid sheet at the equatorial region is shown in Figures \ref{fig:merging}. The images are presented at a time interval of 35 $\mu$s. The ligaments are interconnected with each other during the perforation of the sheet. The ligaments merge with each other leading to a web of liquid in the perforated sheet. The merging and breakup of the liquid ligaments then lead to the formation of spherical droplets. 

\subsection{Fragmentation at the rim of equatorial region}
\label{Fragmentation}
\begin{figure*}
\centering
\includegraphics[width=1\textwidth]{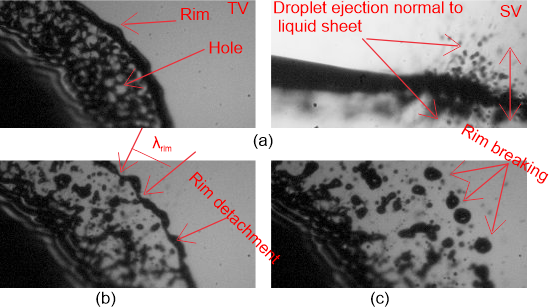}
\caption{\label{fig:breakup} Fragmentation of the liquid sheet: (a) top view (TV) and side view (SV) images during perforation, (b) rim detachment and (c) rim breakup. The corresponding video can be seen in the caption of figure \ref{fig:1}.}
\end{figure*}

The liquid rim detaches from the liquid sheet and breaks into droplets (Figure \ref{fig:breakup} (c)). Unlike the droplets generated by the Faraday instability, which are ejected upwards, perpendicular to the expanding sheet, the droplets generated by rim breakup are ejected and scattered in the direction of sheet expansion. The generation of tiny droplets by the Faraday instability and their ejection in the vertical direction continues until the bag formation starts. The measured size of the tiny-sized droplets, generated by the Faraday instability, varies between 20 $\mu$m to 34 $\mu$m. The wavelength of the deformed rim is around 330 $\mu$m  and breaks into droplets with the size of the order of 100 $\mu$m. At this stage, the generation of the capillary waves of constant wavelength appears on the surface of the liquid sheet. The capillary wavelength on the surface of the liquid sheet is 65 $\mu$m. A detailed discussion on the droplet size distribution is presented later.  \\

\begin{figure*}
\centering
\includegraphics[width=1\textwidth]{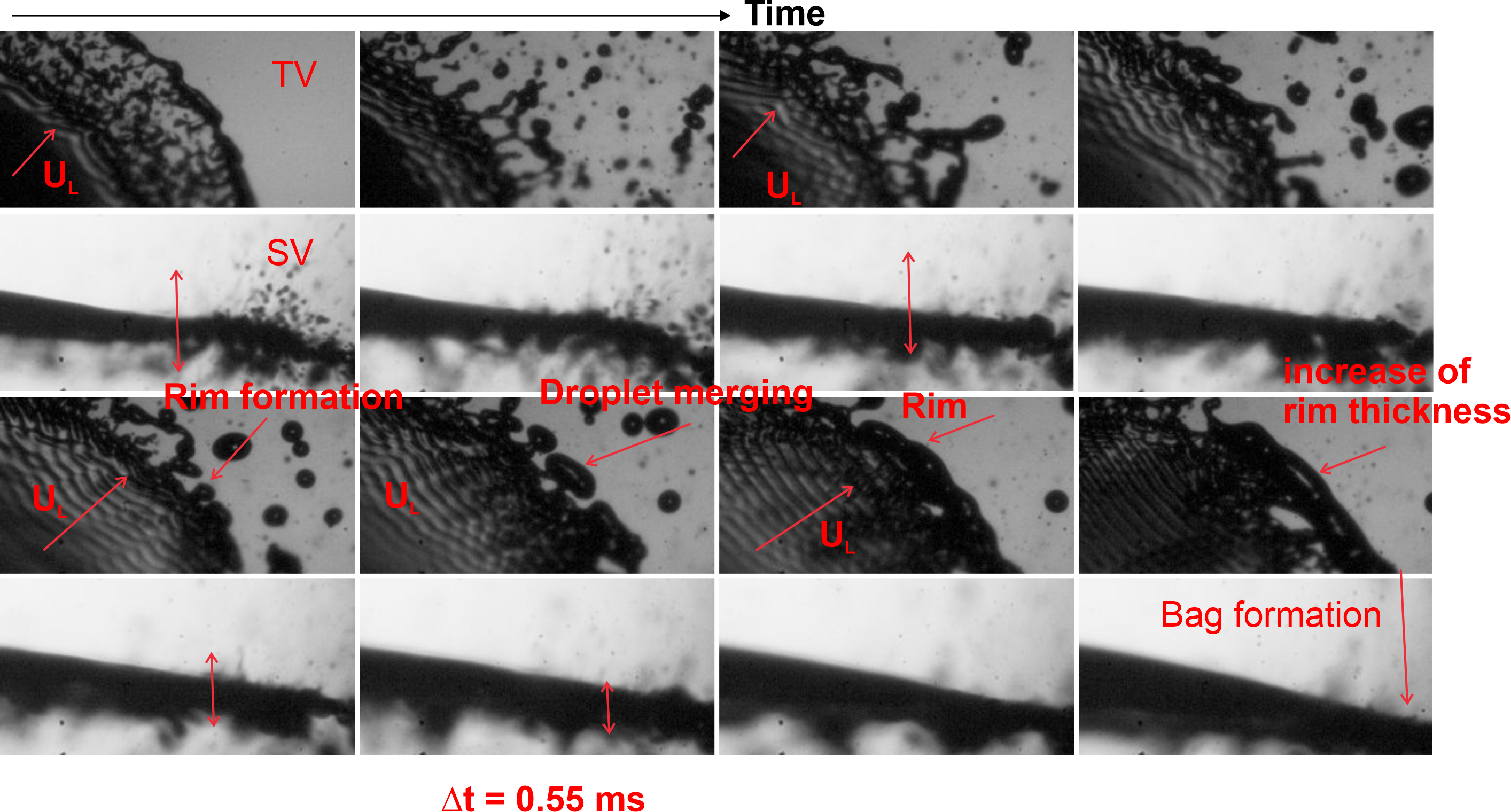}
\caption{\label{fig:bagformation} Instantaneous top and side view digital images during fragmentation at the equatorial region which converted into bag like structure.}
\end{figure*}

Fragmentation at the equatorial region of the droplet continues during the radial expansion of the decelerating liquid sheet as shown in Figure \ref{fig:bagformation}. During this period, the droplets are ejected in both radial and perpendicular to the liquid sheet.  The liquid sheet, after undergoing a complete breakup of the thinning region at the edge of the sheet near the equator, leads to the cessation of further growth and marks the end of the fragmentation process. The atomisation at the equatorial region strongly depends upon the We (diameter of the droplet). Table \ref{table:wevsteb} shows the equatorial breakup time ($t_{eb}$) as a function of We. Equatorial breakup time decreases with the increase of We (an increase in diameter), indicating that the acoustic field hastens the dynamics of the system.

\begin{table}
\caption{\label{table:wevsteb} The deformation time (stretching+thinning) (t), equatorial breakup time ($t_{eb}$) and secondary breakup ($t_{sb}$) as a function of We. Note: $t_sb$ represents the time interval between the end of equatorial atomization and the start of the secondary breakup.}
\begin{center}

\begin{tabular}{ |c|c| c|c|  } 

 \hline
 We &t (ms)& $t_{eb}$ (ms) & $t_{sb}$ (ms)  \\ 
 \hline
 1.37 &7.26 & 4.52 & 2.42 \\ 
 1.45&6.28 & 3.7&3.22 \\ 
 1.56 &5.138&2.6& 5.48 \\ 
 1.92 &1.9& 1.45&15\\ 
 \hline
\end{tabular}
\end{center}
\end{table}

This marks the end of the primary breakup, we term the equatorial breakup of the liquid sheet, whereafter the secondary breakup starts.

\subsection{Bag formation and breakup of different kinds}

The acoustic levitator has the capacity to levitate the droplet only half of its wavelength i.e $\lambda/2 = 4.4 mm$. The sheet diameter is larger than half of the wavelength therefore, the pressure difference between the center and edge of the liquid sheet causes the sheet to bend resulting in the formation of the bag-like shape (see the last image in which the arrow indicates downward). \\

While the droplet breakup at the equatorial region is similar for all $We$, the secondary breakup (i.e. breakup of the bag-like structure) depends upon the diameter of the initial diameter of the droplet, and thereby on the $We$. The observed secondary breakups are classified as (a) umbrella breakup, (b) bag breakup, (c) bubble breakup and (d) rim collapse and sheet breakup (multistage breakup).

\subsubsection{Umbrella breakup}
\label{Umbrella breakup}

\begin{figure*}
\centering
\includegraphics[width=1\textwidth]{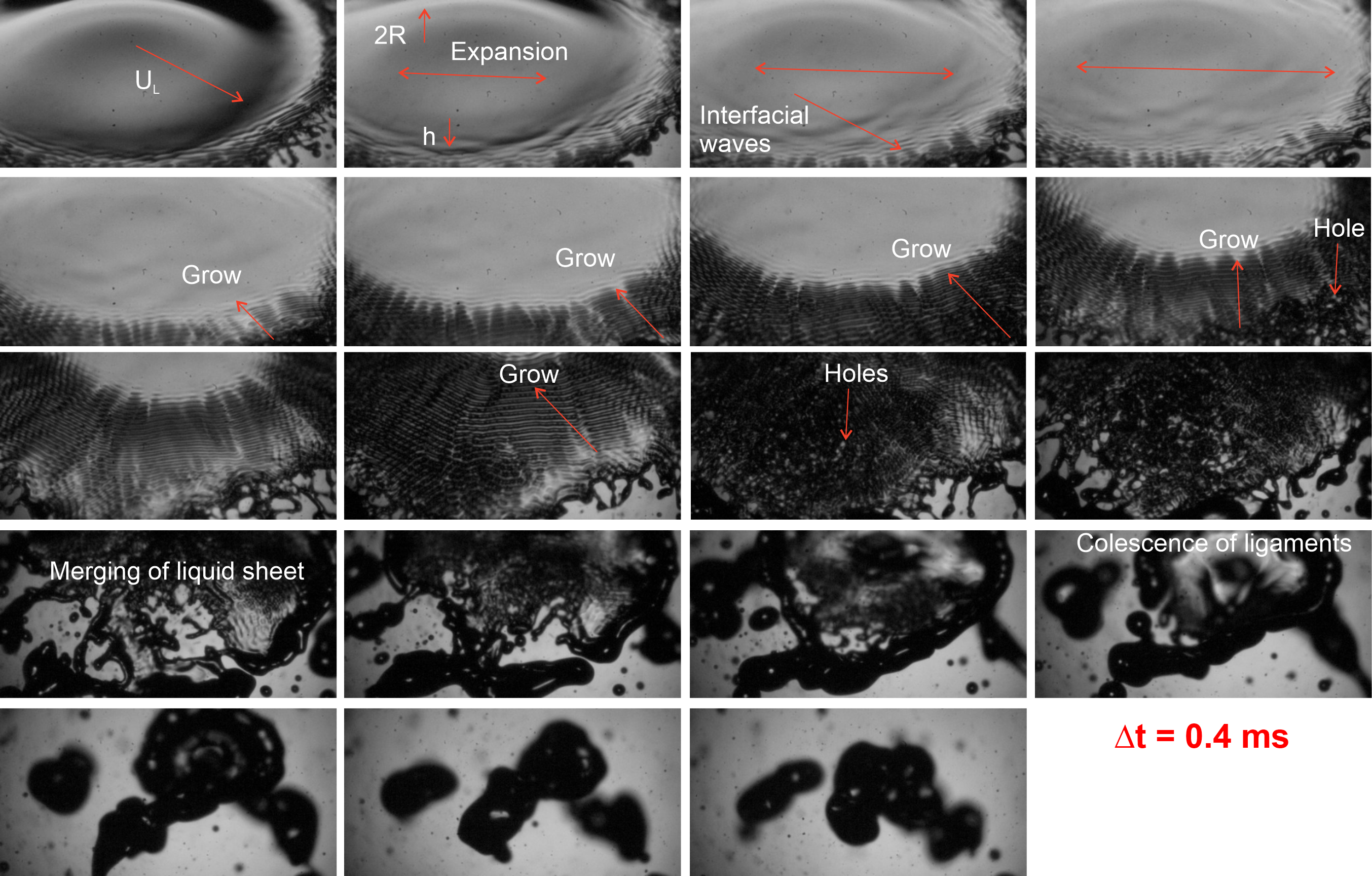}
\caption{\label{fig:umbrella} Instantaneous top view digital images during umbrella breakup regime. The corresponding video is available at \color{blue}  \href{https://drive.google.com/file/d/1Al2M1PRz1Fge2LIbpYGa6qPp3KMy3Gja/view?usp=share_link}{video:Umbrella breakup}}
\end{figure*}

Figure \ref{fig:umbrella} shows the instantaneous digital images of the umbrella breakup of the liquid sheet. The images are recorded at 48k frames per second and presented at a time interval of 0.4 ms. Once a droplet is levitated in the acoustic trap it undergoes stretching, thinning, and Faraday and equatorial breakup, as well as fragmentation and retraction, as discussed in the earlier section \ref{Fragmentationedge} and section \ref{Fragmentation} (The reader is advised to see the details in the movie the link for which is provided in the supplementary material), the thickness of the droplet starts to decrease again as seen in the first picture in figure \ref{fig:umbrella}, and a new thinning region is again formed. The newly formed thinned region shows capillary waves and edge breakup through a combination of Faraday instability, correspondingly formed holes, Rayleigh Taylor, and Rayleigh Plateau mechanisms constituting the secondary breakup mechanism. The liquid droplet that has got transformed into a sheet now thins over the entire region of the droplet. Thinning of the liquid sheet leads to an increase in its frequency to reach the harmonic condition. The harmonic condition leads to the Faraday instability developing on the entire surface of the liquid sheet. The detailed calculation is  discussed in section  \ref{umbmechanism}.\\

\begin{figure*}
\centering
\includegraphics[width=1\textwidth]{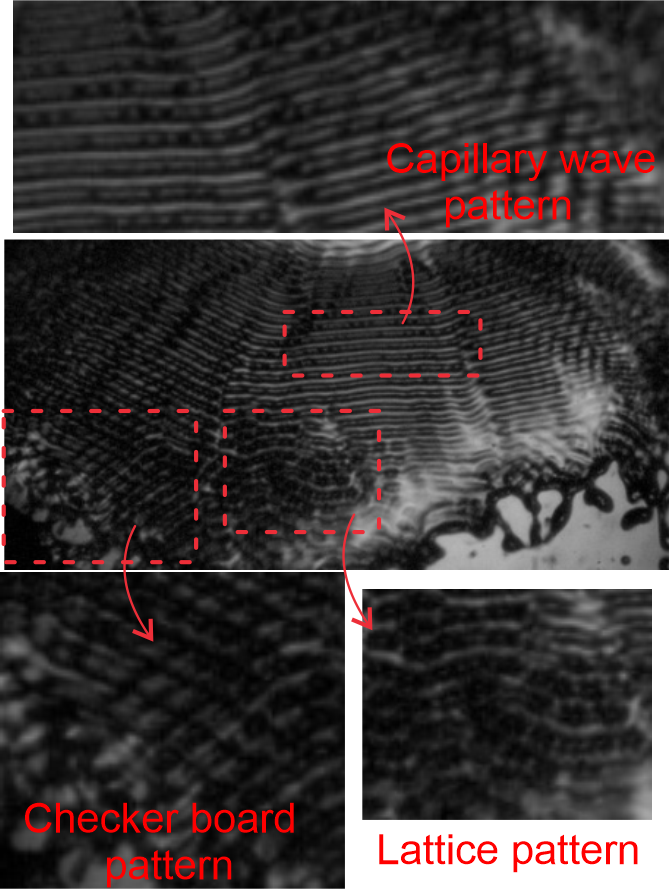}
\caption{\label{fig:pattern} The local variation of the Faraday pattern on the liquid sheet during Umbrella breakup. }
\end{figure*}

 Figure \ref{fig:pattern} shows the development of different types of interfacial instabilities on the liquid sheet during umbrella breakup mode. The capillary wave pattern is developed when the frequency of the liquid sheet becomes equal to half of the applied acoustic frequency due to thinning. The lattice mode and checkerboard pattern are observed at the edge of the sheet during the breakup. The lattice mode pattern was also observed by  \cite{vukasinovic2007dynamics} at the interface of a vertically vibrating sessile droplet and has been attributed to the Faraday instability. The checkerboard pattern is associated with the coupling of the capillary wave pattern with the azimuthal wave. A similar kind of Faraday wave pattern has also been observed elsewhere \cite{shats2014turbulence,khan2019experimental}. \\

The Faraday instability causes the generation and growth of the instability from the equatorial to the center region of the liquid sheet. The Faraday waves formed on the entire surface of the droplet become unstable, causing droplet ejection from the entire liquid sheet and corresponding hole formation. The holes expand and cause the merging of the liquid sheet and result in ligaments that ultimately break due to the RP Instability. \\

The above is termed as umbrella breakup occurs since the sheet, after fragmentation, and before the onset of secondary instability builds a positive, concave upwards curvature.  The liquid sheet completely breaks during the formation of the upward curvature.

\subsubsection{Bag breakup}
\label{bagbreakup}
\begin{figure*}
\centering
\includegraphics[width=1\textwidth]{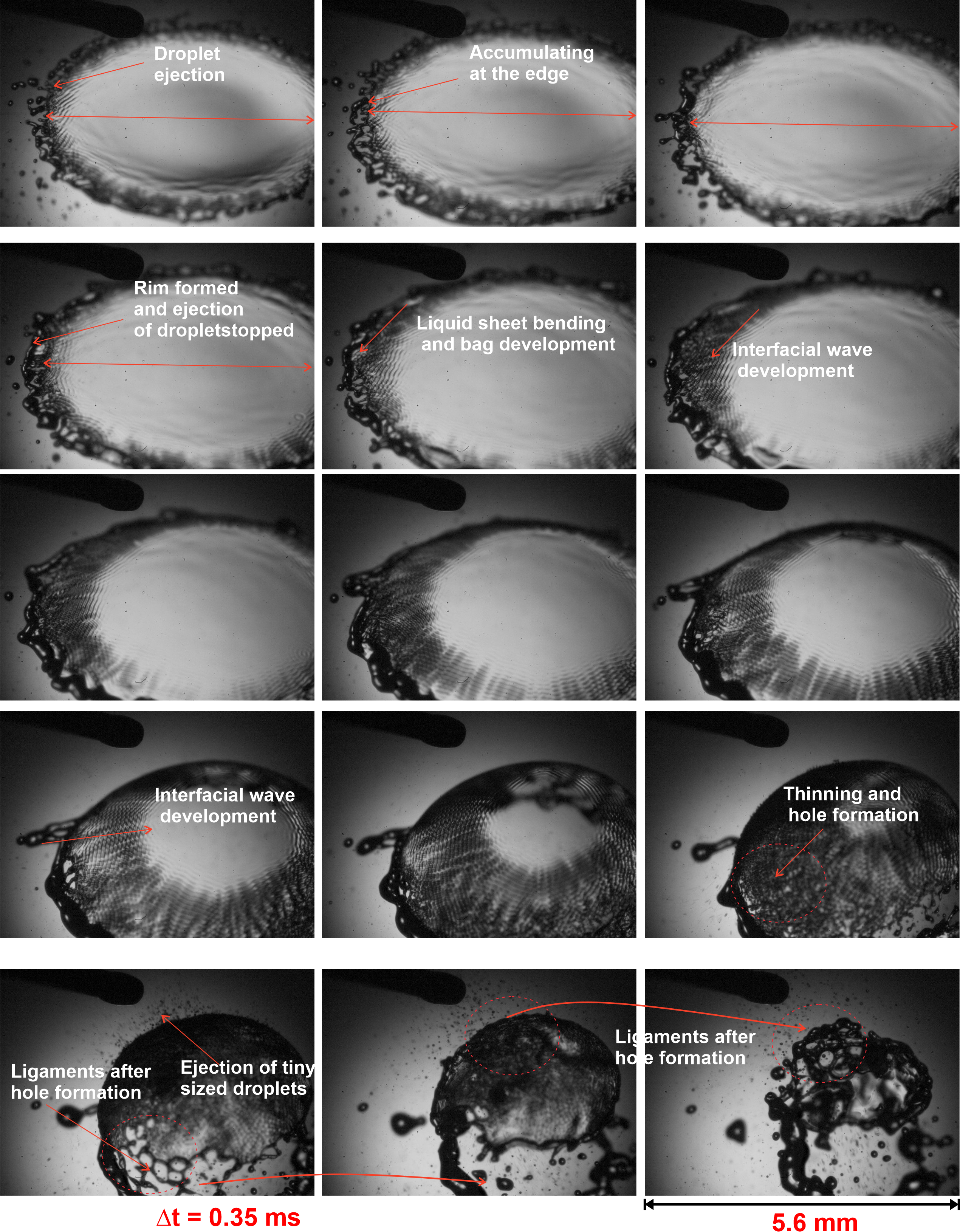}
\caption{\label{fig:bag} Instantaneous top view digital images during bag breakup regime.  The corresponding video is available at {\color{blue}(\href{https://drive.google.com/file/d/1UHPNtUgfoY0brxz5QL6-_DGL77pdqBPl/view?usp=share_link}{video:bag breakup})}.}
\end{figure*}

 Umbrella breakup occurs when the sheet undergoes further thinning and admittance of Faraday instability subsequent to droplet levitation, stretching, thinning, primary equatorial instability, and sheet retraction. On the other hand, bag formation occurs when after the equatorial instability and sheet retraction, the thickness at the equatorial region is not reduced to any significant extent such that harmonic condition and, thereby, the Faraday instability cannot set in. As a result, as the liquid sheet decelerates (decreased rate of expansion of liquid sheet), droplets accumulate at the edge, resulting in the formation of the rim, without any equatorial breakup. The diameter of the liquid sheet can then increase without much further equatorial breakup such that it is beyond the spatial range of the acoustic pressure field in the trap (that varies up to 4.5 mm).  Figure \ref{fig:bag} shows the top view of the digital images recorded during bag formation at 14485 frames per second and displayed at 0.35 ms intervals. The unsupported sheet at the edges bends downwards under gravity and results in the formation of a bag-like structure. The development of the bag increases the surface area of the liquid sheet and the pressure conditions inside the bag lead to significant stretching and, thereby, thinning of the bag. The harmonic condition is thereby achieved by a further reduction in the thickness of the liquid sheet during bag growth. During this time, the Faraday instability is developed at the  surface of the thinned sheet. This causes the droplets to be ejected perpendicular to the sheet, resulting in sheet perforation. The remaining sheet merges and breaks into the larger-sized droplet due to RP instability.

\subsubsection{Bubble breakup}
\label{Bubble breakup}
\begin{figure*}
\centering
\includegraphics[width=1\textwidth]{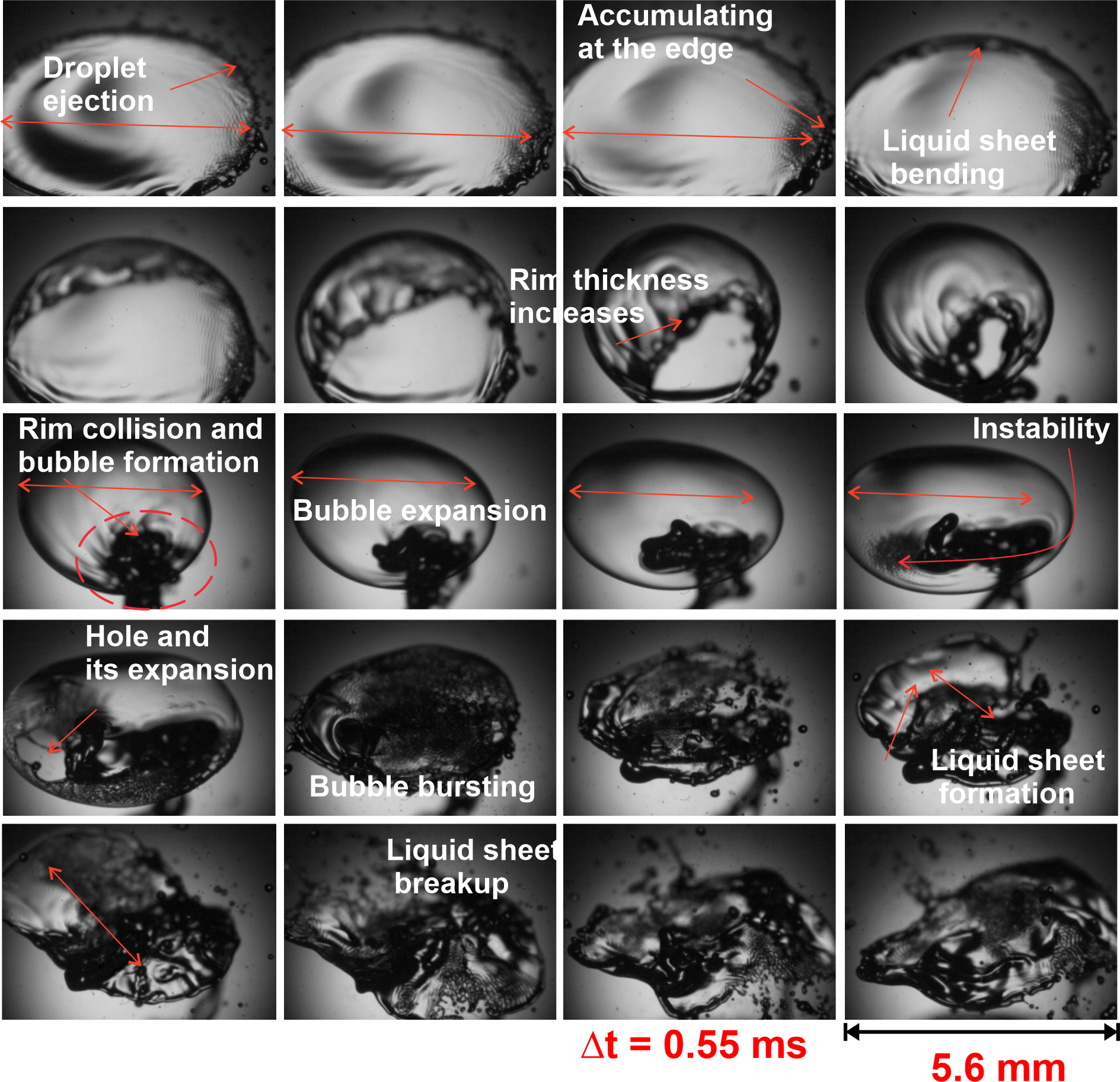}
\caption{\label{fig:bubble} Instantaneous top view digital images during bubble breakup regime. The corresponding video available at  {\color{blue} (\href{https://drive.google.com/file/d/1WlxrkZ1t6fyBqOC6Ma9Cw-gbYaBezq1s/view?usp=share_link} {video:Bubble breakup})}.}
\end{figure*}

Subsequent to droplet levitation, stretching, thinning, and primary equatorial instability followed by sheet retraction, bag formation can occur wherein the bag thins and breaks by Faraday instability. If the thinning of the bag is delayed, the bag (stretched liquid sheet) can close onto itself, resulting in bubble formation. This bubble can then undergo bubble breakup.
Figure \ref{fig:bubble} shows the bubble breakup of a droplet injected in the acoustic levitator. The liquid sheet transforms into a bag-like structure similar to the one discussed in section \ref{bagbreakup}. During bag formation, the thickness of the rim increases with time due to the flow of liquid accumulating at the rim. During rim collision, the bag transforms into a bubble shape shown in Figure \ref{fig:bubble}. Due to the larger thickness of this bubble, no instability is observed at the bubble surface. The bubble then begins to expand into the radial direction with a simultaneous reduction in the thickness of the liquid sheet. The instability emerges at the thinnest region near the rim, as shown in Figure \ref{fig:bubble}. As the thickness further reduces to the magnitude that invokes the Faraday instability, the instability propagates across the entire surface of the bubble resulting in droplet ejection from the surface of the liquid sheet, and a hole is seen to form. The perforated sheet transforms into ligaments and breaks due to RP instability. The breakup of the bubble leads to a residual liquid sheet that again gets levitated the rim deforms and breaks once again. The corresponding video is provided in the caption of the figure \ref{fig:bubble}.

\subsubsection{Multi stage breakup}

\begin{figure*}
\centering
\includegraphics[width=1\textwidth]{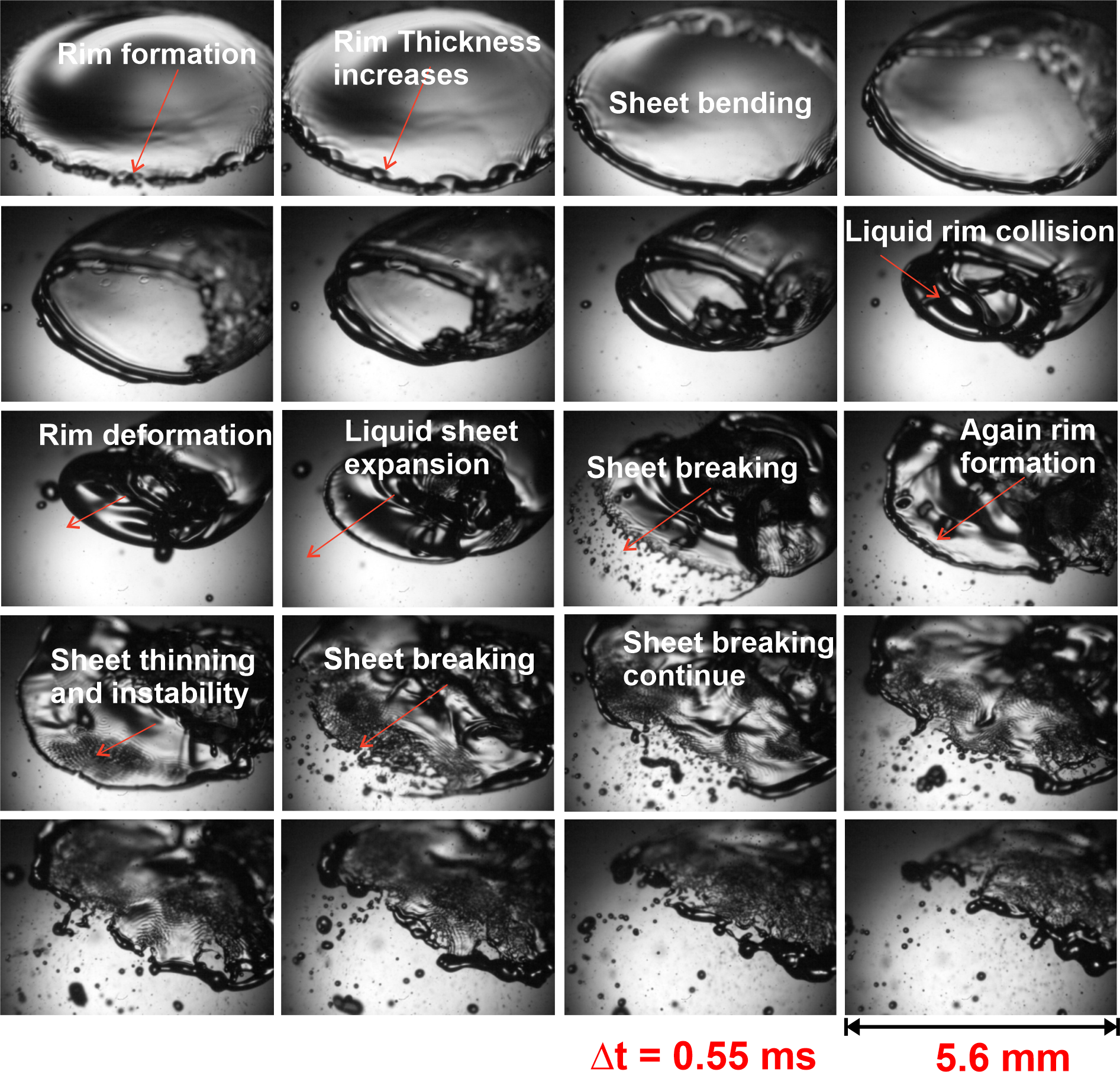}
\caption{\label{fig:multistage} Instantaneous top view digital images during multistage breakup regime. The corresponding video is available at {\color{blue}(\href{https://drive.google.com/file/d/1ZYoEsi2pu7SDWBFynl_1Gk8SkNhp6Lko/view?usp=share_link} {video:multistage breakup})}}
\end{figure*}

Figure \ref{fig:multistage} shows instantaneous images of the multistage sheet breakup. The phenomenon prior to bubble formation (up to rim collision) is similar to the section \ref{Bubble breakup}. Unlike the bubble breakup case, the bubble in the multi-stage breakup does not burst. The rim begins to deform, which leads to equatorial thinning of the sheet.  The surface of this thinned-out bubble is locally similar to a sheet and exhibits fragmentation through a combination of Faraday, RT, and RP instabilities \ref{Fragmentationedge}. The atomization of the sheet is stopped, and a rim is formed once again at the edge of the sheet, such that the sheet again expands and thins locally (see Figure \ref{fig:multistage}) resulting in another cycle of sheet breakup. This breakup continues until a rim forms and the sheet collapses, resulting in ligament formation and breakdown as a result of RP instability. The reader is suggested to refer to the movie provided in the supplementary material for more information.

\section{Droplet size distribution}\label{Dropsize}
As seen from the experimental results presented in the previous section, a variety of instabilities, Faraday, KH, RP and RT, lead to sheet (formed from the original droplet by thinning) breakup. Since each instability is associated with its own unstable wavelengths, together with non-linearities leads to a very complex size distribution of droplets resulting from the atomization. This section includes the size distribution of droplets generated during different regimes of the breakup. The droplet size was measured using Image J software. The estimated error is considered as $\pm 5 \%$. 

\subsection{Droplet size distribution during equatorial breakup}
\begin{figure*}
\centering
\includegraphics[width=1\textwidth]{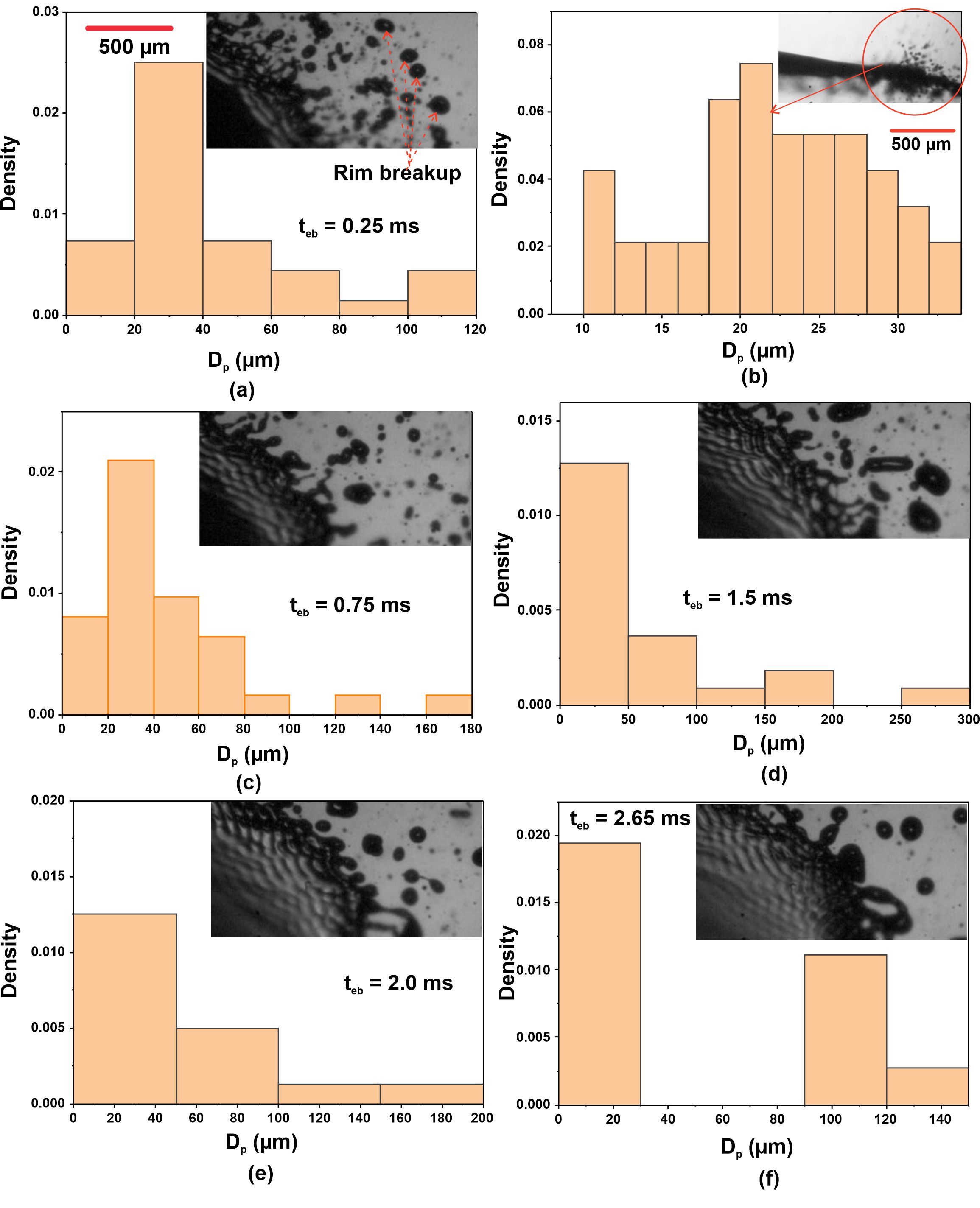}
\caption{\label{fig:sizedistributionequitorial} Droplet size distribution at the equatorial region at the different instant of breakup time. The total breakup time is approximately 2.8 ms. The corresponding visualization video for the equatorial breakup can be found at:{ \color{blue} \href{https://drive.google.com/file/d/1PAOW_pYStmZUSsFsBEUkxwHnGeajwpoB/view?usp=share_link}{video top view: Droplets size equatorial}. \href{https://drive.google.com/file/d/1-C7-TRbNa62WaMq1UCo6olsPJIjY3pp7/view?usp=share_link} {video side view: Faraday droplets}}}
\end{figure*}

Figure \ref{fig:sizedistributionequitorial} shows the size distribution of the droplets generated in the radial direction and normal direction during atomization at the equator of the expanding liquid sheet. The total breakup time at the equatorial region ($t_{eb}$) depends on the We as discussed in an earlier section and for the droplet breakup presented in figure \ref{fig:sizedistributionequitorial}, this is around 2.8 ms. The equatorial breakup time ($t_{eb}$ is the time from the start ($t'=0$) to the end of the atomization process at the equatorial region ($t'=t_{eb}$) and therefore the size distribution of the generated droplets varies with time between $t'=0$ to $t'$. The droplet breakup mechanism consists of (i) Faraday breakup, (ii) RT breakup, and (iii) Capillary breakup (RP instability). The droplet generation perpendicular to the sheet is due to the Faraday instability. On the other hand, the vertical oscillation of the sheet creates radial acceleration that causes RT instability. Figure \ref{fig:sizedistributionequitorial} (a) shows the droplet size distribution at $t'$=0.25 ms. The droplet size distribution here includes rim breakup (indicated by arrow), the capillary breakup of merging ligaments of a perforated liquid sheet, and small-sized droplets generated due to the Faraday instability. These are in fact well captured in figure \ref{fig:merging}. The size of the droplets due to Faraday instability is mostly in the range of 20 to 34 $\mu$m. The size of the droplets generated from the rim breakup varies from 100 to 120 $\mu$m. The droplets generated due to the merging of thin ligaments of perforated sheet vary from 40 to 80 $\mu$m. The contribution of the RT instability is not significant up to this point. Figure \ref{fig:sizedistributionequitorial} (b) shows the droplet size distribution generated by Faraday instability perpendicular to the liquid sheet during the equatorial breakup. The number of generated droplets measured is around 50. The droplet size distribution is found to be in the range of  20 to 34 $\mu$m.

Figure \ref{fig:sizedistributionequitorial} (c) shows the droplet size distribution at $t'$=0.75 ms. For the image at $t'=0.75ms$, most of the droplets generated at $t'=0.25ms$, would have gone out of the frame such that the image mostly has droplets that are generated between the time period $t'=0.25-0.75$ ms. As the rim separates from the thin liquid sheet, the thin liquid sheet vibrates vertically and generates a radial acceleration, resulting in the radial generation of droplets due to RT instability. The mechanism of formation of radially generated droplets due to RT instability will be discussed in the next section. The size of the droplets varies from 60 to 80 $\mu$m. The smaller-sized droplets (20 to 34 $\mu$m) in this distribution correspond to the Faraday instability, which continues to be at play. \\

\begin{figure*}
\centering
\includegraphics[width=1\textwidth]{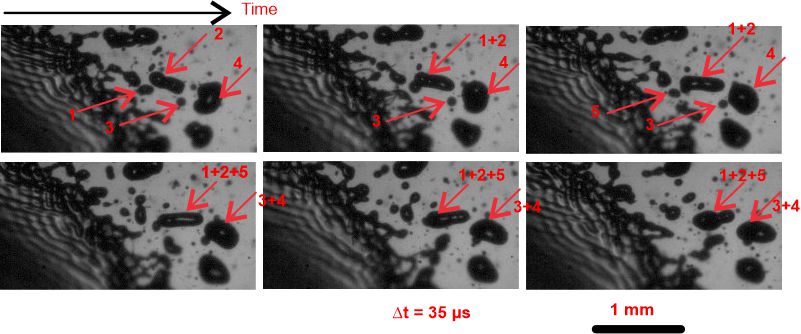}
\caption{\label{fig:mergingtwodroplets} Merging of two droplets in radial direction.}
\end{figure*}

 The RT instability is suppressed as the liquid velocity decreases with time (deceleration of the liquid sheet). In this regime ($t'=0.75-1.5 ms$) the breakup takes place because of combined RT and capillary instability as shown in Figure  \ref{fig:sizedistributionequitorial} (d) to yield droplets of size 100 to 200 $\mu$m. The larger-sized droplet in Figure \ref{fig:sizedistributionequitorial} (d) is because of the merging of the droplets (see Figure \ref{fig:mergingtwodroplets}). \\

The equatorial breakup continues with a weakening of the RT instability with time (see  \ref{fig:sizedistributionequitorial} (e)). The process of merging droplets is significantly suppressed here, whereby larger-sized droplets ($>$ 200 $\mu$m) are not formed ($t'=1.5-2.0 ms$). \\

\begin{figure*}
\centering
\includegraphics[width=1\textwidth]{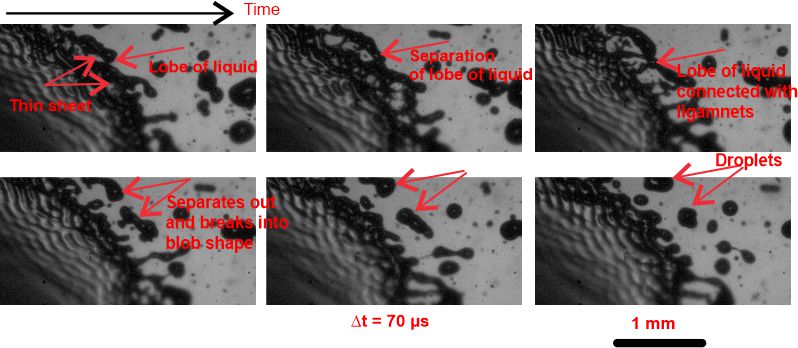}
\caption{\label{fig:capillarybreakup } Droplet generation at the late stage of the equatorial breakup (deceleration phase) during smaller radial velocity}
\end{figure*}

The capillary breakup takes over the RT instability between $t'$= 2 ms and $t'$= 2.65 ms which can be clearly seen in Figure \ref{fig:sizedistributionequitorial} (f). The deceleration of the sheet and relative acceleration leads to the suppression of RT instability. Moreover, the thinning region reduces in extent. The prevailing vertical generation of the droplet causes perforation and separation of the edge rim of the liquid sheet to create liquid threads and ligaments which break into droplets by capillary breakup Figure \ref{fig:capillarybreakup}. The Faraday instability too seems to have reduced at this point. The reader is suggested to refer to the movie provided in the supplementary material for more information.


\subsection{Droplet size distribution of very tiny sized droplet ejected perpendicular to liquid sheet and larger sized during bag breakup}
\begin{figure*}
\centering
\includegraphics[width=1\textwidth]{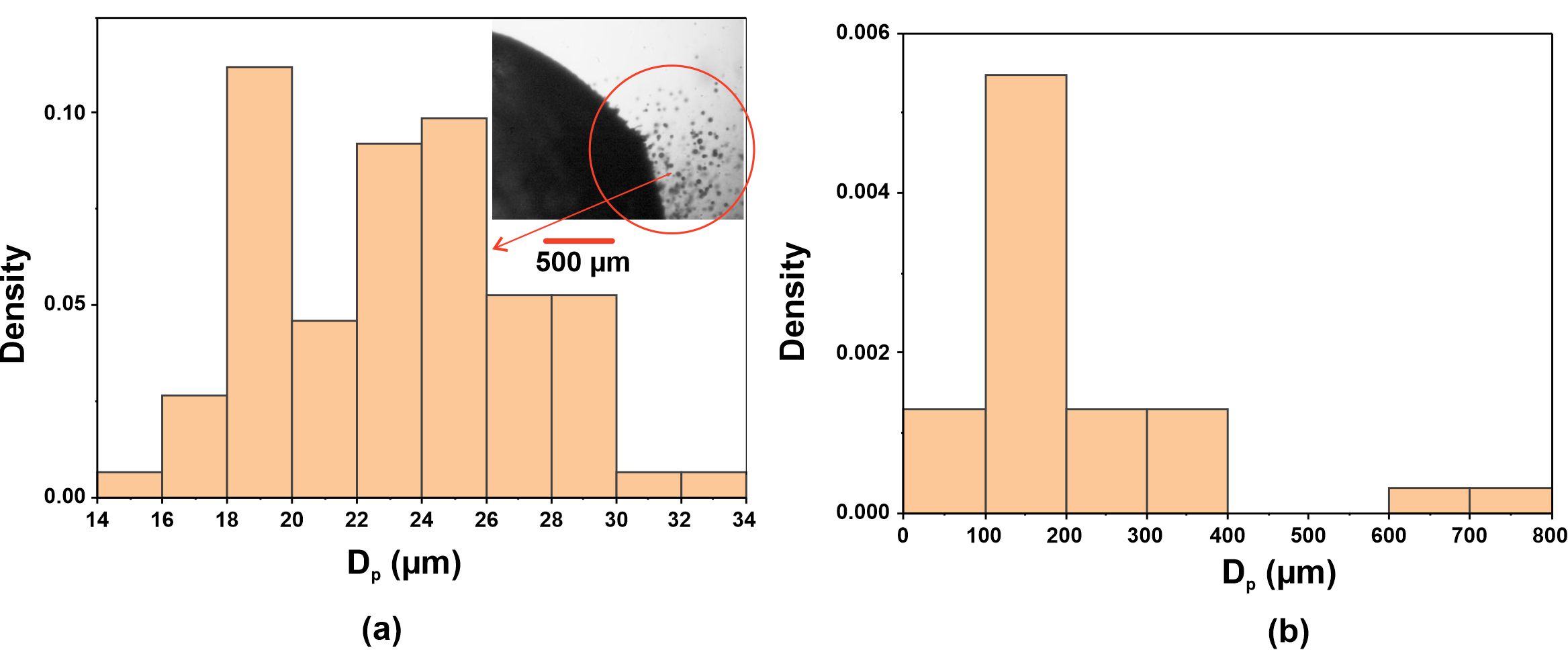}
\caption{\label{fig:bagdroplets} Droplet size distribution of droplet ejected during bag development: (a) tiny droplets (faraday breakup) and (b) larger droplets (after sheet perforation due to RP instability). The corresponding visualisation video can be found at{\color{blue} \href{https://drive.google.com/file/d/1fCwPrESqWA8fc_NfNJZfn1SVmnbC-Kjy/view?usp=share_link}{video:Bag Faraday}}.}
\end{figure*}

Figure \ref{fig:bagdroplets} (a) shows the droplet size distribution of tiny droplets produced from the Faraday instability during bag formation. Droplet size distribution is measured over 10 consecutive images captured at 48000 fps (measured from figure \ref{fig:dropletejection}). The sizes of the droplets are found to vary in the range of 20 to 34 $\mu$m, similar to those formed during equatorial breakup thereby confirming their formation by the Faraday instability. Figure \ref{fig:bagdroplets} (b) shows the droplet size distribution of the generated droplets during bag breakup after the sheet perforation. The size of the droplets varies between 100 to 200 $\mu$ m, confirming that the droplets break due to RP instability (rather than RT instability). 
 Much larger droplets to the tune of 0.3mm to 0.7mm are formed by the liquid rim breakup as well as the merging of ligaments unperforated liquid sheets, which collapse and form ligaments. These droplets have a larger diameter of 0.3 mm to 0.7 mm. This is clearly demonstrated in the video link provided in the figure caption. This phenomenon occurs not only for complete bag breakup, but also for other regimes such as umbrella, bubble, and multi-stage. Therefore, it can be concluded that the large droplets are always a result of liquid rim breakup and the breakup of ligaments formed by the merging of the liquid sheet. 

\section{Discussion}
The droplet breakup in swirl flow or gravitational fall as well as breakup of sheets formed by co-axial jets, have been investigated in numerous studies. Similarly, droplet oscillation, levitation, and break up in an acoustic levitator have been addressed to some extent in the literature.  In this section, we invoke the results of these studies to understand the mechanism of droplet levitation, stretching, thinning, equatorial breakup, bag formation, interfacial instabilities, and droplet size distributions in our experimental observations detailed in sections \ref{experiments} and \ref{Dropsize} on acoustically levitated droplets. 
\subsection{Mechanism of Droplet Levitation, deformation and critical We}
In appendix \ref{Derivation of pressure variation} we show that the total acoustic radiation force on the drop, which when balanced by the weight of the droplet, yields the droplet position in the trap below the node. The calculations show that a 2 mm droplet is levitated around 300 microns below the center node position of the standing wave. Given that the node-to-node distance is around 4.5 mm, the droplet can be considered to be located almost exactly at the center of the node of the standing wave. \\

The initial shape of the droplet is assumed to be spherical. The spherical droplet changes to an oblate cylindrical shape in the standing wave. Therefore, h can be expressed as

\begin{equation}
    h \sim \frac{d_0^3}{6 R^2}
\end{equation}
where $d_0$ is the diameter of the undeformed droplet.  The final shape of the droplet is due to the inertial pressure of the gas and the surface tension force. 
\begin{equation}
    2 \pi R ( 2 \sigma) = \rho_a U^2 \times 2 \pi R h
\end{equation}

By eliminating $h$ in the above equation
\begin{equation}
\label{staticR}
    \frac{2R}{d_0} = \frac{1}{\sqrt{3}} We^{1/2}
\end{equation}
Thus, the theoretical results from equation \ref{staticR} indicate that the size of the deformed droplet  $\frac{2R}{d_0}$ $\simeq$ $We^{1/2}$ in agreement with experiments. 

The droplet then breaks at a critical We = 1.36. \cite{danilov1992breakup} argue that the restoring force due to the curvature of the elipsoidal droplet at the equator cannot balance the deforming negative pressure at the equator, whereby the droplet cannot attain a steady shape. They predict a critical We of around 1.7.

\subsection{Mechanism of Flattening}
\begin{figure*}
\centering
\includegraphics[width=0.8\textwidth]{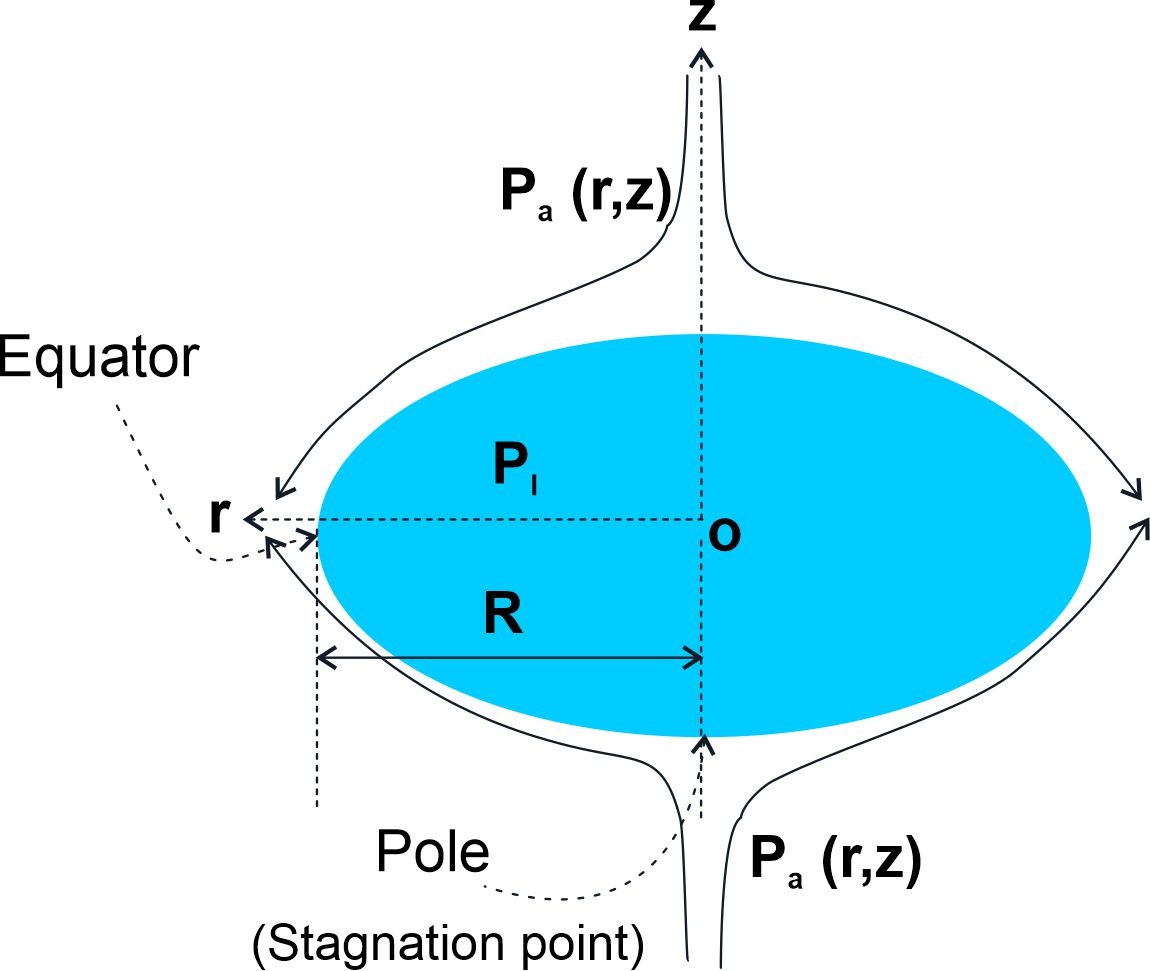}
\caption{\label{fig:flattenmechanism} Sketch of the mechanism of droplet flattening. }
\end{figure*}

For a droplet levitated at the node of a stationary sound wave, the pressure field is approximated by \ref{Dvelocity} as provided in the appendix. 
With the assumption of inviscid, incompressible fluid flow, the current model determines droplet deformation in accordance with the mechanisms in references  \cite{villermaux2009single,kulkarni2014bag,kirar2022experimental}. Thus for a droplet at the node of a standing wave in an acoustic field, the droplet is subjected to a time-varying pressure and velocity field, which in turn yields a  non-zero time averaged pressure field, on account of the Bernoulli equation, as given by,
\begin{equation}
    P_a (r,z) = P_a (0) - \rho \frac{f^2 U^2}{8d_0^2} r^2 - \rho \frac{f^2 U^2}{2d_0^2} z^2
\end{equation}
Where, $r = \sqrt{x^2+y^2}$ is the radial coordinate and $P_a = \frac{\rho_a U^2}{2}$ is the stagnation pressure at z = 0 (see Figure \ref{fig:flattenmechanism}). The air velocity, $ U = \frac{P_0}{\rho_0 v_0}  \frac{1}{\sqrt{2}}$, the details about the calculation of U are given in the \ref{Dvelocity}. The stretch factor is highly shape-dependent and is expected to get nonlinear with $We$. Here, as in other literature too, we use $f$ as a fitting parameter.

 The Navier-Stokes equation and continuity equation have been solved to predict the deformation of the droplet in the stretching regime. The experimental observation is illustrated in Figure \ref{fig:stages} for We = 1.45. 

The Navier-Stokes equation and continuity equation in radial coordinates are given as,
\begin{equation}
    \rho_l \Bigg(\frac{\partial u_r}{\partial t}+ u_r \frac{\partial u_r}{\partial r}\Bigg)= \frac{-\partial p}{\partial r}+\mu_L \Bigg[\frac{1}{r} \frac{\partial}{\partial r} \Bigg(r \frac{\partial u}{\partial r}\Bigg)-\frac{u_r}{r^2}\Bigg]
\end{equation}

\begin{equation}
    r \frac{\partial h}{\partial t}+ u_r \frac{ \partial (r u_r h)}{\partial r} = 0
\end{equation}
The change in the height of the droplet during radial expansion is  written as, $h = \frac{d_0^3}{6R^2}$ and the velocity field inside the droplet is written as,
\begin{equation}
u (r,t) =  \frac{r}{R} \frac{dR}{dt}
\end{equation}
The deformation of the droplet is due to the pressure difference between the liquid and surrounding pressure. By neglecting the tangential component, the normal component of the stress is balanced as, 
\begin{equation}
    \frac{2 \sigma_l}{h} = P_L (R) - P_a (R)
\end{equation}
 $\frac{2 \sigma_l}{h}$ is the capillary pressure at at the liquid-air interface. $P_L(R)$ and $P_a(R)=P_a(R,0)$ are the inside (liquid) and outside pressures (surrounding) respectively. By assuming the origin of a cylindrical coordinate system at the center of the droplet (z=0), the pressure at the edge of the droplet, $r = R$ is given below.

\begin{equation}
    P_L (R) = P_a (0) - \rho \frac{f^2 U^2}{8d_0^2} R^2 + \frac{2 \sigma_L}{h}
\end{equation}
 Defining $\phi = \frac{2R}{d_0}$,  T = $t/\tau$, $\tau= \frac{d_0}{U} \sqrt{\frac{\rho_L}{\rho_a}}$. Note as discussed in the appendix, $P_a(0)=P_a(0,h)$. Thus the evolution of the diameter $R$ of the deformed droplet can be given as,
 
 \begin{equation}
 \label{eqn:flat1}
     \frac{d^2\phi}{dT^2}-\Bigg(\frac{f^2}{4}- \frac{24}{We} \Bigg) = 0
 \end{equation}
 
 The solution of the above equation by using the boundary condition at $\phi (0) = 1$ and $\phi' (0) = 0$  is given as,
 \begin{equation}
 \label{phi}
 \phi(T) = \frac{1}{2} exp\bigg(-\frac{\alpha T}{2}\bigg) \bigg(1+ exp\big(\alpha T\big)\bigg)
 \end{equation}
 
where,  $ \alpha= \frac{\sqrt{-96+f^2 We}}{\sqrt{We}}$

\begin{figure*}
\centering
\includegraphics[width=1\textwidth]{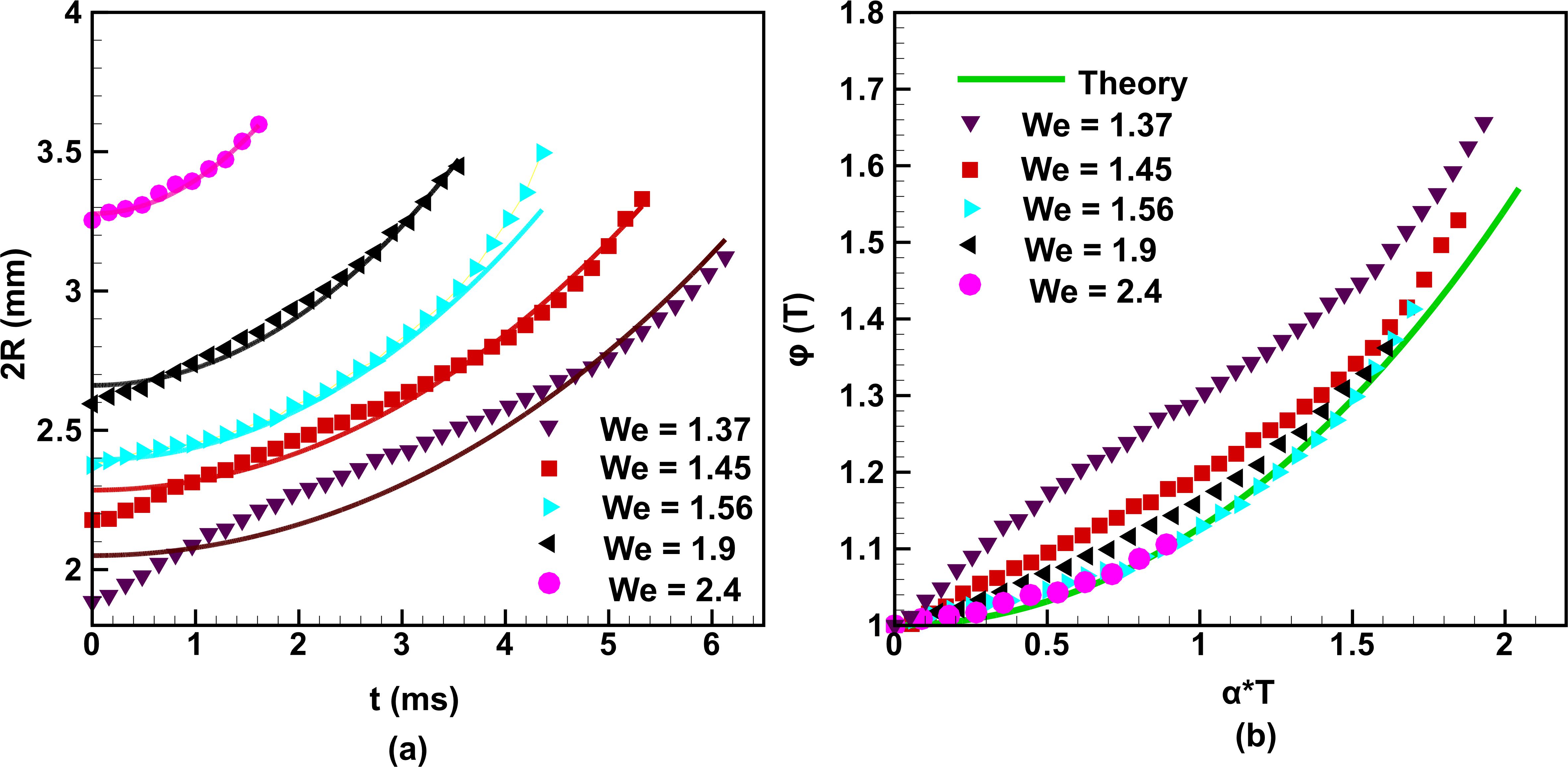}
\caption{\label{fig:phivsT} The comparison of experimental and theoretical results in the stretching regime: (a) 2R vs t (b) $\phi$ vs $\alpha$*T .}
\end{figure*}

The solid line and symbol in figure \ref{fig:phivsT} (a) represent the evolution of 2R with t obtained from the equation \ref{phi} and experiments, respectively.  Figure \ref{fig:phivsT} (b) shows a good data collapse between experiment and theory and the data collapses is seen to be better for higher Weber number. The difference between the experiments and theory at smaller We may be attributed to the oscillations of the smaller-sized droplet in the standing wave. It is important to note that the stretching factor (f) varies with droplet size (or We) as listed in the appendix.

It is useful to see if equation \ref{eqn:flat1}, which describes stretching of the droplet with time, as a function of $We$, can be used to estimate the critical $We$. At low We, the stretching factor is found to asymptote to a value of $f=9.2$. The critical We beyond which a droplet is always stable, and shows stretching can now be estimated using equation \ref{eqn:flat1}, assuming a steady state, such that $We_c=96/f^2$. This yields $We_c \sim 1.14$ in close agreement with the experimental value of around 1.36 reported in the results section.

\subsection{Mechanism of thinning}
\begin{figure*}
\centering
\includegraphics[width=1\textwidth]{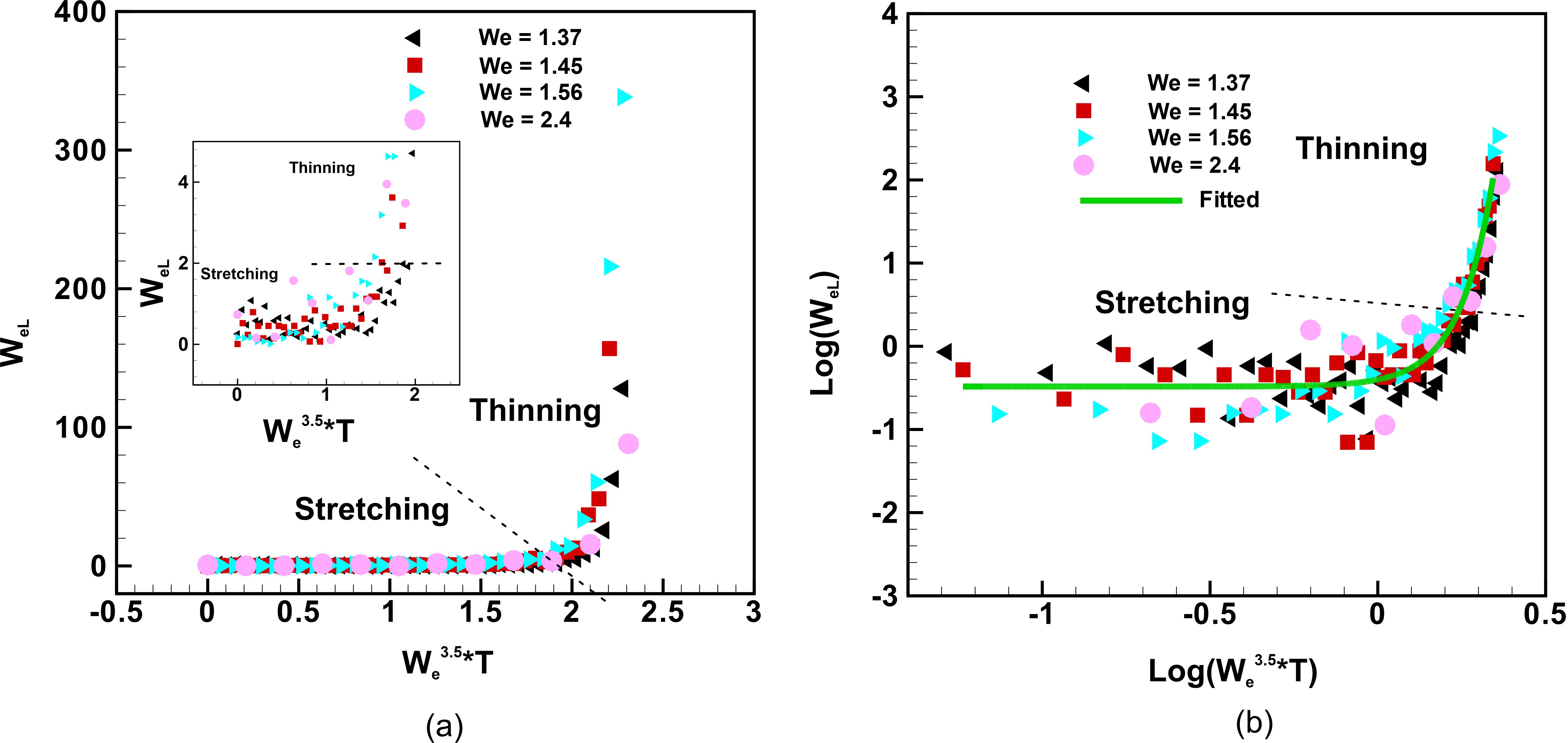}
\caption{\label{fig:WeL} (a) $W_{eL}$ vs $We^{3.5} *T$ and (b) $Log(W_{eL})$ vs $Log(We^{3.5} *T)$.}
\end{figure*}

 The droplet placed in the acoustic standing wave expands radially with a reduction of the height (thickness of the droplet) of the droplet. This thickness is approximately uniform along the radial extent during the stretching regime. The radial velocity of the expanding liquid is already presented in section \ref{experiments}. The radial velocity can be expressed in terms of a non-dimensional liquid Weber number defined as $W_{eL}$ = $\frac{\rho_L U_L^2 d_o}{\sigma}$. It was discussed in section \ref{section:droplet thinning}, that the stretching regime is followed by the thinning regime, where towards the equatorial edge, a thin sheet is formed at the edge of the stretched droplet. The radial expansion in this regime is highly nonlinear. It is then attempted to explain both the stretching and the thinning regime with a single empirical expression. Figure \ref{fig:WeL} shows the variation of the $W_{eL}$ with respect to scaled time and the radial diameter during liquid sheet expansion. The $W_{eL}$ changes slowly in the stretching regime (see Figure \ref{fig:WeL}(a) and (b)). A critical  $W_{eL}$ is obtained as 2 which can be clearly observed in zoomed view inserted in Figure\ref{fig:WeL}(a), beyond which the onset of stretching can be observed. Thus the non-dimensional variation of the radial extent of the sheet with time can be explained by a single empirical expression, $Log(We_L/0.6)=0.08 \; exp[10\; Log(We^{3.5}*T)]$. 

 Numerical calculations were conducted by inserting a droplet at the node with an aspect ratio chosen in accordance with the results of the experiment. The results indicated that in the stretching regime, the pressure difference between the pole and the equator varies as $P_P - P_E$ $\simeq$ $R^{1/2}$ and $P_P - P_E$ $\simeq$ $R^{11/5}$ (see in appendix figure \ref{fig:comsol} (h)). This illustrates that the thinning of the equatorial region is triggered by the sudden rise in the $P_P - P_E$. 

 \subsection{Mechanism of Primary breakup: atomization at equatorial region}

\subsubsection{Fragmentation at the thinned equatorial region before rim breakup: Faraday instability}
The Faraday waves are nonlinear standing waves that develop on the surface of a vibrating liquid when the frequency increases above a certain threshold. These waves are known to exhibit a variety of patterns and ripples on the surface of the vibrating liquid. The response frequency of these waves could be sub-harmonic or harmonic with respect to the imposed acoustic frequency. The levitator used in the present study has an imposed acoustic frequency of 40k Hz. 

The frequency of the waves on a liquid surface can be expressed as (\cite{kumar1994parametric,bush2015pilot,khan2019experimental}).
\begin{equation}
\label{oscfrequency}
    \omega_0^2=(g k +\frac{\sigma k^3}{\rho_L}) Tanh(kh)
\end{equation}
where the first term corresponds to gravity waves and the second to capillary waves. 
The smaller wavelengths seen in the present study imply that the $\frac{\sigma k^3}{\rho_L}$ $>>$ gk, thereby rendering the gravitational effects negligible.
The calculated value of the $\omega_o = 127.1 krad/s$, correspondingly $f_o=20.23 kHz$ for the experimental $\lambda = 65 \mu m$ ($k = \frac{ 2 \pi}{\lambda}$). The ratio of the $\frac{\omega_0}{\omega} \sim  0.51$, which shows the liquid sheet frequency is half of the applied frequency $\omega$ and is thus in the sub-harmonic region. This is in good agreement with the experimentally observed oscillation frequency of the sheet, measured at 140,000 fps, which yields a value of $20$ kHz. It is known that a liquid sheet shows parametric resonance only when the thickness of the liquid sheet reduces to such an extent that $\omega_0$ is half the forcing frequency. For higher thicknesses, $\omega_0>\omega/2$, where $\omega$ is the frequency of the sound waves, which explains why only the thinned part of the stretched droplet undergoes Faraday instability.

\subsubsection{KH Instability}

The Kelvin-Helmholtz type inter-facial waves develop when a thin liquid sheet radially expands in the air. The dispersion relation for the KH instability of a moving liquid sheet of constant thickness was first investigated by \cite{squire1953investigation}. This dispersion relation is valid only for the long wave approximation i.e. $kh\ll 1$ developed two possible modes of KH waves symmetric (varicose) and asymmetric (sinuous mode). The  dispersion relation for a sinuous mode of the liquid sheet with constant thickness h moving with a velocity $U_L$, which is excited by vertical vibration, is written as (\cite{bremond2007atomization})
\begin{equation}
\label{KH1}
    \frac{\omega}{kU}= \frac{1\pm (\alpha coth(kh/2))^{1/2} [-1+\frac{\sigma k}{\rho_l U_L^2} (1+\alpha coth(kh/2)]^{1/2}}{1+ \alpha coth(kh/2)}
\end{equation}
Where, $k = \frac{2 \pi}{\lambda}$ is the wave number, $\lambda$ is the wavelength and $\alpha = \frac{\rho_a}{\rho_l}$ is the ratio of densities.

The equation \ref{KH1} is valid for the constant thickness which is not true for real situations because thickness decreases with R as confirmed by the observation. Therefore, Bremond et al \cite{bremond2007atomization} modified the above equation for variable thickness $h=\frac{d_0^2}{8R}$. The dispersion relation in non-dimensional form is given as
\begin{equation}
\label{KH2}
\tilde{\omega} = \frac{\tilde{k}^2}{\tilde{k}+16\alpha \tilde{R}} \Biggl(1\pm\Bigg[\frac{16 \tilde{R}}{We_l}\bigg(1+\frac{16\alpha \tilde{R}}{\tilde{k}}\bigg)-\frac{16\alpha \tilde{R}}{\tilde{k}}\Bigg]^{1/2}\Biggl)
\end{equation}
Where, $\tilde{k} = k d_0$, $\tilde{R} = R/d_0$, $We_l = \frac{\rho_L U_L^2 d_0}{\sigma}$, $\sigma$ is the surface tension and $d_0$ is the initial diameter of the droplet.
The pulsation is equal to the real part of the above equation as the frequency of the acoustic field is equal to the sinuous frequency. Therefore, the real part is expressed as 
\begin{equation}
\label{KH3}
    \tilde{k}(\tilde{R}) = \frac{\tilde{\omega}_0}{2} \Bigg[1+\Bigg(1+\frac{64 \alpha \tilde{R}}{\tilde{\omega_0}}\Bigg)^{1/2}\Bigg]
\end{equation}

Here, $\tilde{w}_0 = 2 \pi f d_0 /U_L$ is the frequency of the acoustic field. For $f = 40000 Hz$, $d_0 =$ 2 mm, $U_L = 2.5 m/s$, $\rho_L = 830 kg/m^3$, $\rho_a = 1.22 kg/m^3$, $\tilde{R}=1.37$, the wavelength comes out to be  $\lambda = 63.16 \mu m$. Our experimental results show that the wavelength on the surface of the liquid sheet in the thin region is equal to 65 $\mu m$ which shows good agreement with the wavelength calculated by the equation \ref{KH3}. This confirms that the waves on the surface of the liquid sheet are sinuous modes of KH waves. Now we need to verify if this wave number is stable or unstable at this imposed frequency.  \\ 

The imposed oscillation frequency is sinuous mode with amplification and damped. Thus at a critical radius beyond which the sheet is completely stable and the amplitude of oscillation will be constant (\cite{bremond2007atomization}). We have assumed that the initial diameter of the droplet is equal to the critical radius and cut off frequency has been calculated by the equation 
\begin{equation}
\label{KH4}
    \tilde{\omega}_c = \alpha We_L \Bigg(1-\frac{16 \tilde{R}_c}{We_L} \Bigg)^{1/2}
\end{equation}
The cut-off frequency comes out equal to $ f_0 = 102 Hz$ ($We_L$ = 350). The system is unstable for the frequency smaller than $ f_0$ which is very small compared to the imposed frequency (40k Hz). Therefore, the waves on the surface are not unstable wave and oscillate with constant amplitude and wavelength. Therefore, it is confirmed that the instability on the surface is not due to the sinuous mode of the KH instability. 

The vertical vibration is not visible for the imaging conducted at smaller fps compared to the imposed frequency. This is because the vertical vibration is driven by the applied frequency. 

\begin{figure*}
\centering
\includegraphics[width=1\textwidth]{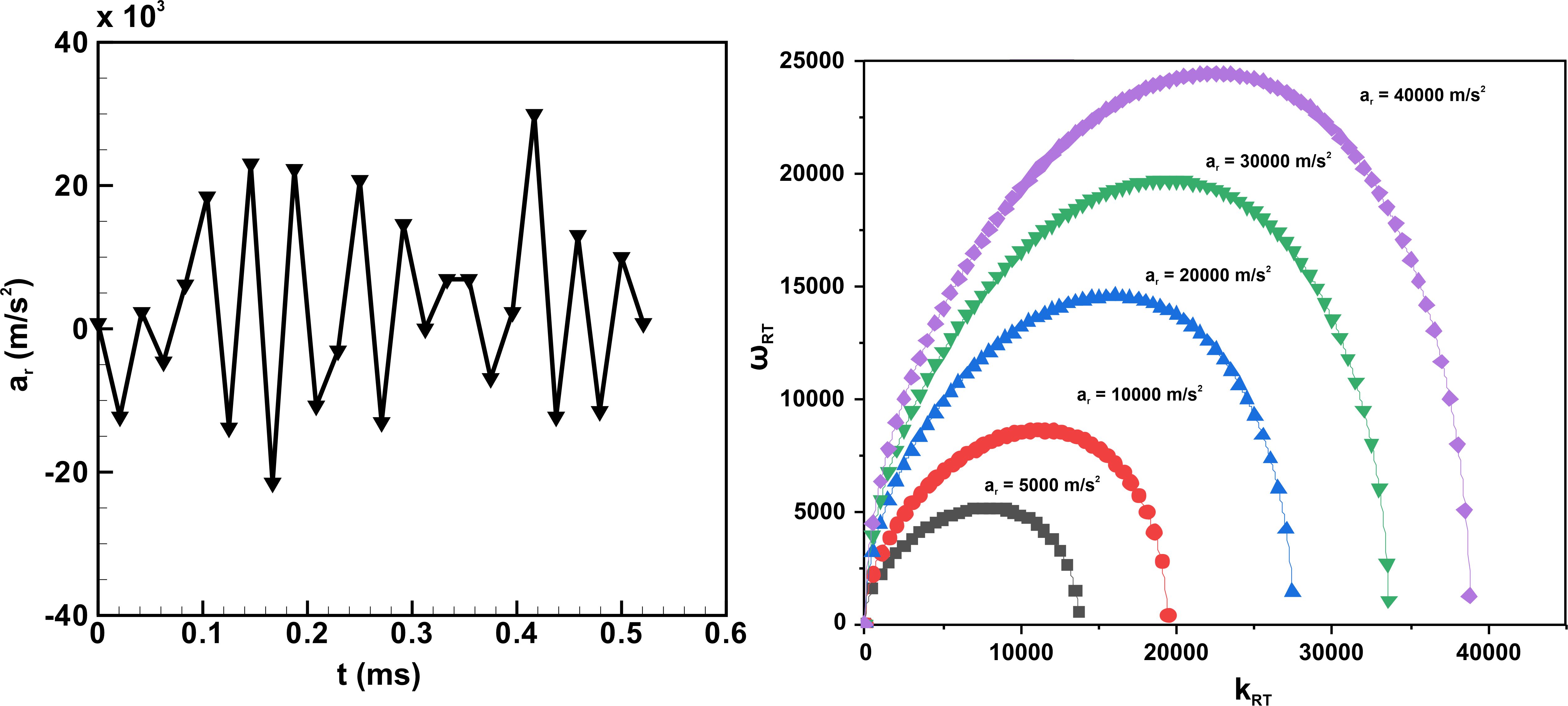}
\caption{\label{fig:ar} (a) Radial acceleration of the rim with time during propagating in the air (thinning regime) before initiation of breakup which is calculated from Figure \ref{fig:velandampl} (a), and (b) RT instability growth rate as a function of wave number.}
\end{figure*}

\subsubsection{Fragmentation at the rim of the equatorial region: Rayleigh Taylor instability}

 The Faraday waves appear only at the thinned region of the sheet preceding the rim of the liquid sheet and eventually when the Faraday instability sets in, it results in the ejection of tiny-sized droplets as presented in the section \ref{Fragmentationedge}. During this process, the thinned part of the liquid sheet vibrates vertically with increasing amplitude. This fluctuation of velocity generates radial oscillatory acceleration that also leads to instantaneous deceleration of the sheet. The calculated acceleration is given in Figure \ref{fig:ar}.

 \begin{figure*}
\centering
\includegraphics[width=1\textwidth]{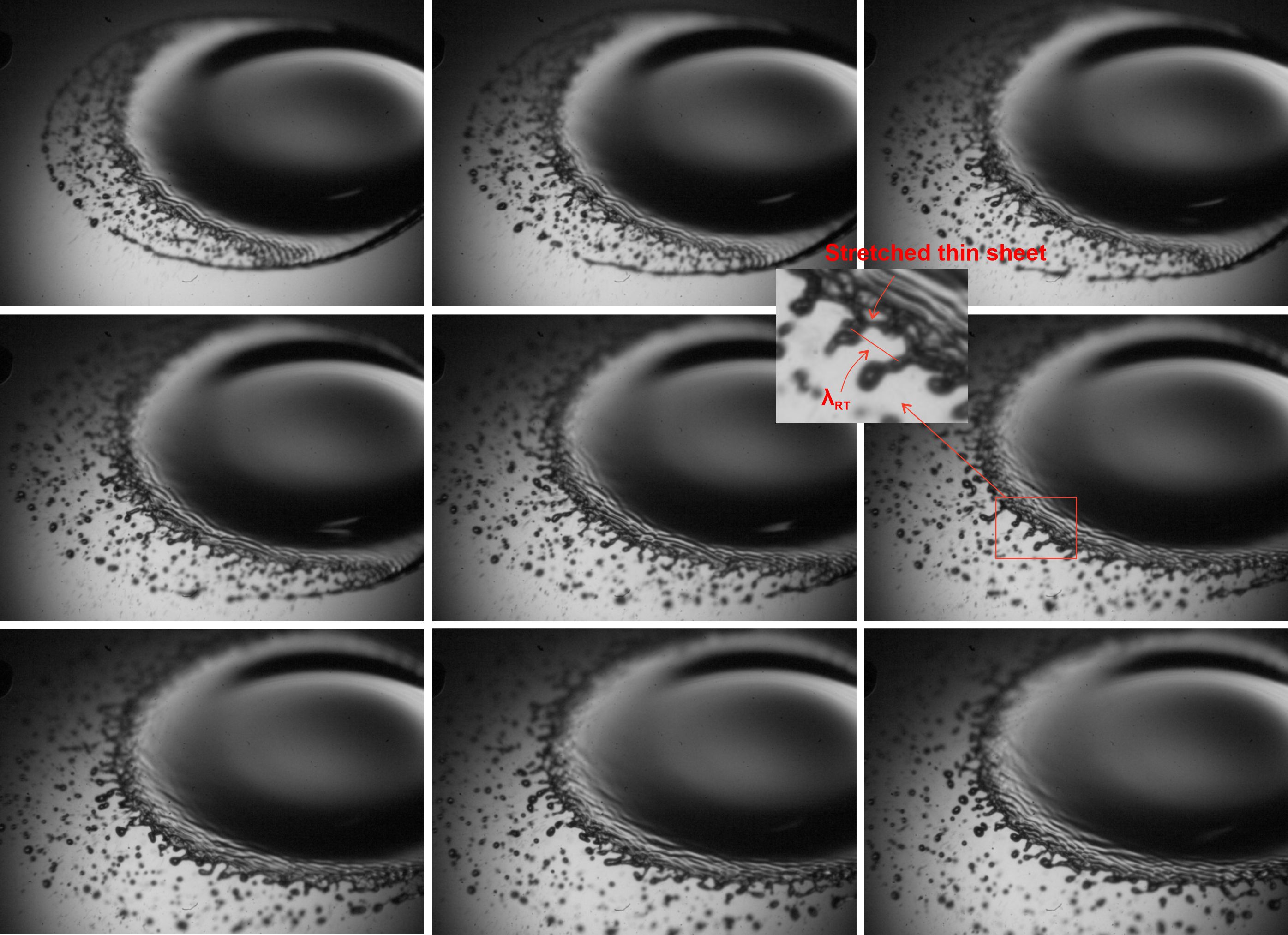}
\caption{\label{fig:RT} Instantaneous top view digital images during breakup at the equatorial region. The images are shown at the time interval $\Delta t = 69 \mu s$. The width of the image is equal to 5.62 mm. The corresponding video can be seen at {\color{blue} \href{https://drive.google.com/file/d/1jMVeUfnjTFtpUToeq7kr3w2okPapSCcz/view?usp=share_link}{video:RT instability}}}
\end{figure*}

 The droplet generation mechanism at the end of the sheet, is due to the formation of holes due to droplet ejection by Faraday instability and the corresponding stream-wise stretching of the perforated thin liquid sheet, which results in the formation of ligaments. These ligaments then break into droplets Figure \ref{fig:RT} (see video provided in the caption) by a Rayleigh Plateau instability. The formation of ligaments and their subsequent breakup is further exacerbated by the Rayleigh Taylor instability. 
 The ligament generation in the radial direction is facilitated by the radial acceleration and deceleration in the perforated sheet that can lead to the Rayleigh Taylor instability (\cite{jarrahbashi2014vorticity}). The dispersion relation for the RT instability and its wavelength can be determined by (\cite{jarrahbashi2014vorticity,sharma2021shock}).
\begin{equation}
    \omega_{RT}= \bigg(\frac{a (\rho_L - \rho_a) k_{RT} - k_{RT}^3 \sigma}{\rho_L  + \rho_a}\bigg)^{1/2}
\end{equation}
where a is the acceleration,  and $\omega_{RT}$ is the dimensional growth rate. The corresponding most unstable wavelength is determined as
\begin{equation}
\label{waveRT}
    \lambda_{RT} =  2 \pi \sqrt{\frac{3\sigma}{a (\rho_L - \rho_a)} }
\end{equation}

The radial acceleration of the oscillating rim during radially propagating in the air is calculated from the variation in velocity Figure \ref{fig:velandampl} (a) and presented in Figure \ref{fig:ar}. The maximum radial acceleration measured on the rim is 29000 $m/s^2$ and the corresponding azimuthal wavelength as measured in experiments is 0.338 mm. The calculated wavelength by the equation \ref{waveRT} is equal to 0.329 mm, indicating a fairly good agreement with the experiments. 

The breakup of the thinned sheet at the equatorial region of the droplet, results in delayed breakup of the rim. This is essential because even before the rim breaks, the Rayleigh Taylor and Faraday instabilities already start generating a significant number of droplets in the highly perforated sheet preceding the rim. Aided by these perforations, the rim ultimately partly separates and undergoes breakup by Rayleigh Plateau instability. \\

It is useful to mention here that even before the sheet breakup starts, the azimuthal deformation observed in the rim before breakup is not observed prior to the vertical oscillation (see Figure \ref{fig:rim} (a)) of the thinned sheet. The thinned sheet preceding the rim, when sufficiently thin, oscillates in a vertical direction while radially propagating in the air (in the thinning regime), which results in the azimuthal deformation on the rim (see Figure  \ref{fig:rim} (c) and (d)) on account of the RT instability. The azimuthal wavelength of the rim compares well with the most unstable wavelength of the RT instability. These azimuthal undulations infact propagate backward and create radially propagating azimuthally distributed undulations. \\

As the sheet keeps ejecting droplets, by Faraday instability in a direction perpendicular to the plane of the sheet, and droplets generated by RT and RP instabilities in the direction of radial expansion of the sheet,  the liquid sheet subsequently breaks into droplets of larger diameter with a decrease in the liquid velocity; the merging of lobes leads to the accumulation of drops that re-forms the rim as the surface tension force overcomes the inertia force.

\subsection{Mechanism of secondary breakup}
\label{umbmechanism}
This section includes the mechanism behind the secondary breakup (after the equatorial breakup). The secondary breakup includes umbrella breakup, bag breakup, bubble breakup, and multi-stage breakup. The mechanism of each will be discussed in the following sections.

\subsubsection{Mechanism of umbrella breakup}
The observation of the umbrella breakup is reported in section \ref{Umbrella breakup}. The umbrella breakup occurs for  1.36 $\le$ We $\le$ 1.45 i.e. for smaller droplets. A typical droplet, after undergoing equatorial breakup, and rim retraction, bends upwards to form a concave shape stretched droplet sheet, resembling an umbrella. With increasing We the curvature of the umbrella increases. Once the umbrella is formed and the umbrella-shaped droplet stretches further to again get into the thinning regime, equatorial breakup and KH as well as Faraday waves obeying the frequency of the vibrating sheet to be half that of the applied frequency, start to appear on the surface of the liquid sheet. \\

To confirm this, we measured the wave velocity by extracting the images frame-by-frame during wave propagation on the surface of the liquid sheet, which was found to be of the order of 2.5 $m/s$. By assuming that this velocity is equal to the liquid velocity, the corresponding thickness is determined by the Taylor-Culick criteria to be equal to 8.5 $\mu$m. The wavelength,  65 $\mu$m, and the corresponding frequency of oscillation, \ref{oscfrequency} agree well with the calculated oscillation frequency ($\omega_0$) of 127.1K, which is half of the imposed acoustic frequency (251.2K). This is a clear indication of sub-harmonic condition. Thus the wave appears on the entire surface of the liquid sheet when the liquid sheet thickness reduces ($U_L$ increases) such that the frequency of the liquid sheet becomes equal to half of the applied acoustic frequency. It should be noted that the capillary wavelength can be estimated as $\frac{U_L}{f} = 62.5 \mu$m, which demonstrates a surprisingly good agreement with the measured wavelength. Once, the capillary wave appears on the surface, the instability pattern changes at different locations of the liquid sheet as shown in figure \ref{fig:pattern}. This may be because of the uneven thinning (locally) of the liquid sheet that causes the development of different modes of Faraday instability. The Faraday instability leads to perforation of the stretched umbrella-like liquid sheet leading to the merging of the ligaments and their breakup due to RP instability.

For better clarity, the capillary wave propagation on the surface of the liquid surface in the sub-harmonic condition can be seen in the video: {\color{blue}\href{https://drive.google.com/file/d/1sN973DT0W7tM8iAbKBHRoQ39F3PhbYpD/view?usp=share_link}{capillary wave formation and instability}}. This top-view video was recorded by focusing on the liquid sheet surface at 140000 fps.

\subsubsection{Mechanism of Bag breakup}
\label{mechanismbag}
The bag formation commences during the deceleration of the liquid sheet towards the end of the equatorial breakup process  (see Figure \ref{fig:bagformation}). In this case, when the $We_{L}$ decreases and the surface tension dominates over the inertia force. Therefore the ligaments and detached liquid rim (due to capillary breakup) do not break into droplets and remain attached to the edge of the liquid sheet. This results in an increase of the sheet thickness (see the video given in the caption of figure \ref{fig:sizedistributionequitorial}) and prevents breakup by umbrella mode (Faraday instabilty). Since the diameter of the stretched liquid sheet is beyond the levitation capacity of the levitator, the edges of this stretched liquid sheet curl downwards in the vertical direction. The pressure distribution on the liquid sheet is such that it is maximum at the pole and minimum at the edge of the sheet (see the appendix section in figure \ref{acoustic thinning}). This leads to the liquid sheet bending like a bag during its simultaneous upward movement due to acoustic radiation pressure.

\begin{figure*}
\centering
\includegraphics[width=1\textwidth]{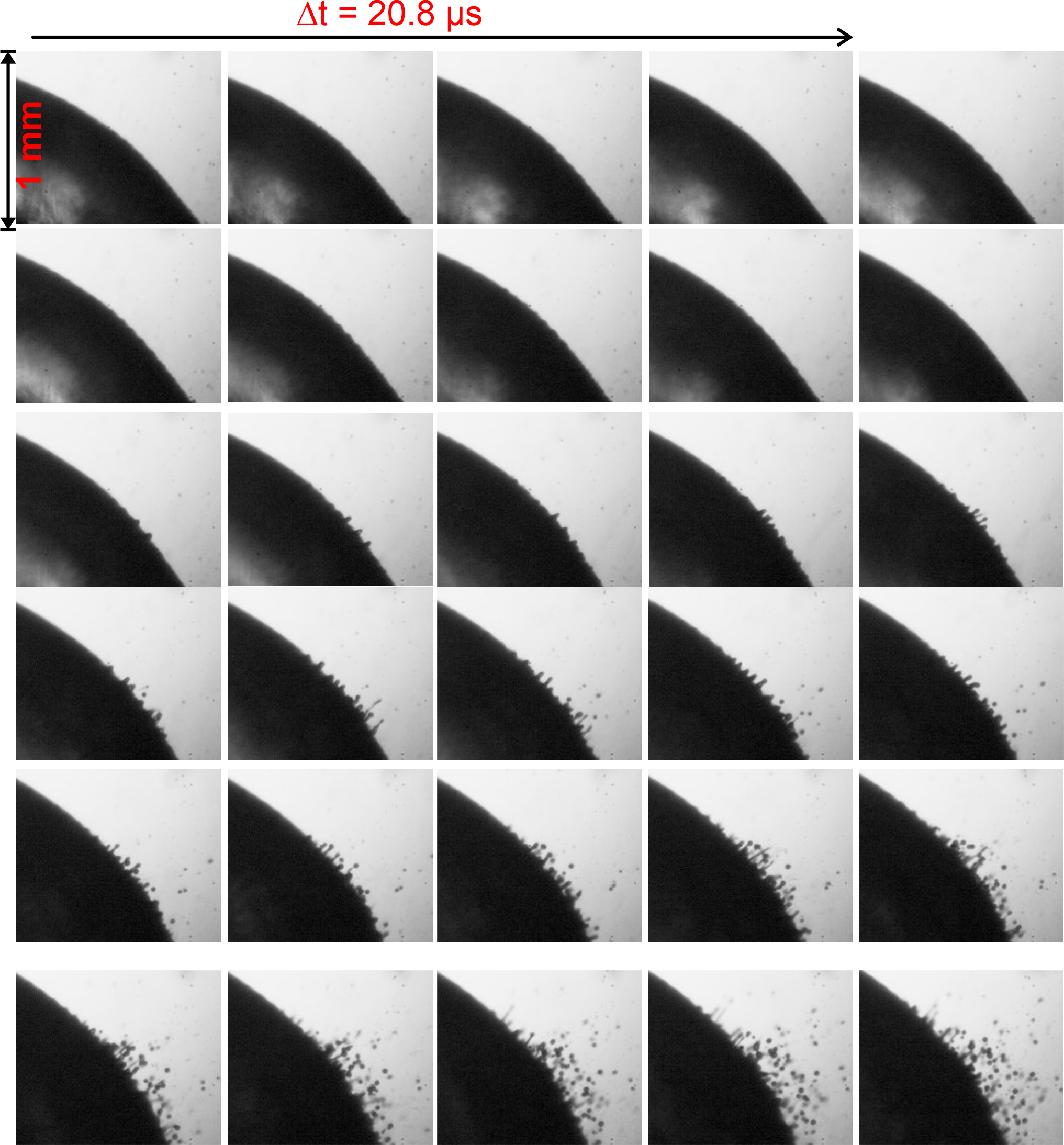}
\caption{\label{fig:dropletejection} Instantaneous side view digital images of droplet ejection during propagation of instability in bag breakup regime. The images are presented at a time interval of 20.8 $\mu$s. The corresponding video available at{\color{blue}\href{https://drive.google.com/file/d/1fCwPrESqWA8fc_NfNJZfn1SVmnbC-Kjy/view?usp=share_link}{video:Bagfaraday}}.}
\end{figure*}

\begin{figure*}
\centering
\includegraphics[width=1 \textwidth]{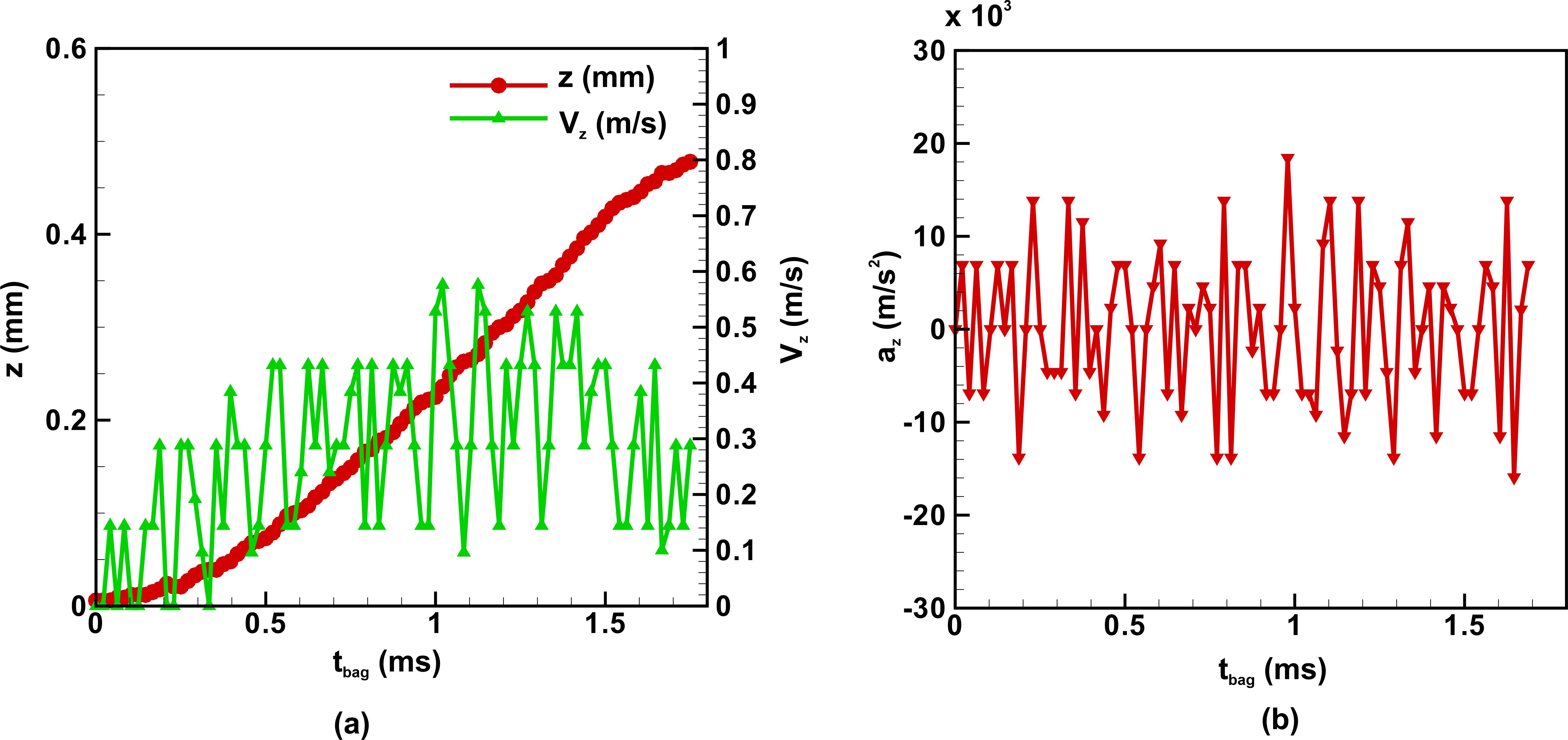}
\caption{\label{fig:vzandaz} The change of height (z), velocity ($v_z$) and acceleration ($a_z$) during bag formation (upward movement of the liquid sheet). The vertical movement is calculated by focusing at the center of the liquid sheet. Here, $t_{bag}$ denotes the time between the end of the equatorial breakup and the start of the secondary breakup.}
\end{figure*}

The extent of upward movement (z) during bag development and subsequent ejection of the tiny-sized droplets, perpendicular to the liquid sheet due to Faraday instability, are shown in figure \ref{fig:dropletejection}. This occurs as the thickness of the liquid sheet decreases to meet the prerequisite for sub-harmonic condition. For additional information, the reader is advised see the video given in the caption of figure \ref{fig:dropletejection}. Figure \ref{fig:vzandaz} shows the distance traveled (z) in the upward direction and the corresponding velocity and acceleration of the liquid sheet during bag growth. It is seen that $z$ increases exponentially with time before the initiation of a hole in the sheet. During this period, the time-averaged velocity increases for a period of 1 ms with increased acceleration and subsequent onset of deceleration where after, it starts decreasing.  The periodic acceleration and deceleration of the liquid sheet in the vertical direction during bag formation ( Figure \ref{fig:vzandaz} (b))   suggests the possibility of RT instability perpendicular to the surface of the liquid sheet. The RT wavelength estimated corresponding to the maximum acceleration/deceleration of the bag (12500 $m/s^2$) is $\lambda_{RT}$ = 0.502 mm. The measured wavelength of the deformed liquid rim during upward propagation of the bag is equal to 0.72 mm (by top view visualization) and 0.6 mm (by side view visualization). The theoretical and measured values agree reasonably well, suggesting that the RT instability may be responsible for the deformation of the bag rim.\\

\begin{figure*}
\centering
\includegraphics[width=1 \textwidth]{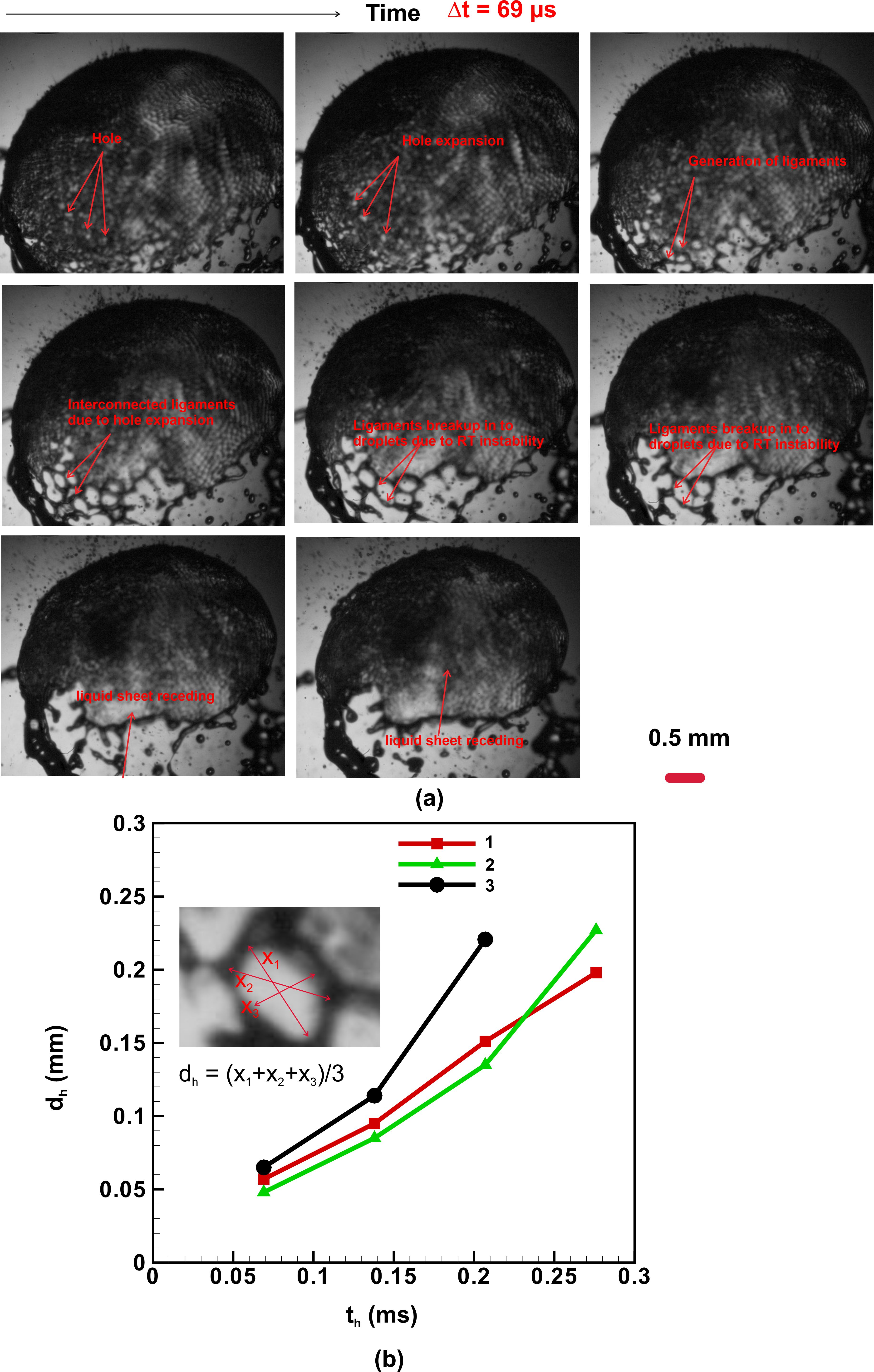}
\caption{\label{fig:holemech} Hole formation and ligaments breakup during bag breakup regime: (a) visualization of top view images and (d) hole expansion diameter with time. Here, $d_h$ is the hole diameter and $t_h$ is the time starting from the hole formation.}
\end{figure*}

The velocity decrease in figure \ref{fig:vzandaz} appears to be after holes form due to the ejection of tiny-sized droplets caused by the Faraday instability of the thinned region of the bag.  This reduction of the thickness plays an important role in secondary breakup and depends upon the We number. The sub-harmonic condition is only set up when the liquid sheet frequency becomes on the order of half of the imposed acoustic frequency. As soon as the thickness reduces to reach the sub-harmonic region the wave appears on the surface and droplet ejection starts as the instability setup. A  video of Faraday instability ({\color{blue} \href{https://drive.google.com/file/d/1ayq3Y5FjI1gG9DTUb9fwouWUQFWzkQb1/view?usp=share_link}{Faraday instability}}) on the surface of the bag is captured at a frame rate of 140k fps. The ejection of the droplets due to Faraday instability creates a hole (see the first image of Figure \ref{fig:holemech}). A few such holes are formed, which expand radially with the formation of the liquid rim at the edge. The hole rim thickness increases during hole expansion. The expansion of the holes creates interconnected ligaments which become unstable due to RP instability and break into the droplets. This process continues until the complete sheet breakup of the liquid sheet. This video provides a nice illustration of how the liquid sheet has changed into thick ligaments during perforation, which then breaks into the larger size droplet: ({\color{blue} \href{https://drive.google.com/file/d/1d-NhcmOLBJOkd0L-9CiPCQIkb4nsCsuP/view?usp=share_link}{liquid sheet to ligaments}}). This video has been recorded at 72 000 fps from top view. To calculate the rate of expansion of the hole with time, we measured the diameter of the hole with time focusing with 3 holes. The change in the diameter of the hole with time is presented in Figure \ref{fig:holemech} (b).

Figure \ref{fig:holemech} (b) shows the expansion of the hole diameter ($d_h$) with time for 3 such holes with typical velocities of the order of $0.4-0.5$m/s. The hole expansion at a constant velocity of a perforated liquid sheet is also reported by \cite{liu2022numerical,fraser1962drop,taylor1959dynamics,culick1960comments}. Contrary to this, some researchers have observed exponential growth \cite{{debregeas1995viscous,roth2005evidence,savva2009viscous}}. Our results also show an onset of exponential growth $d_h$ with time (see Figure \ref{fig:holemech} (b) 2 and 3). In this case the exponential velocity increases with time. The expansion diameter $d_h \sim exp(t/\tau)$ is identical to the observation of \cite{debregeas1995viscous}.  

\subsubsection{Mechanism of Bubble breakup and Multistage breakup}

\begin{figure*}
\centering
\includegraphics[width=1\textwidth]{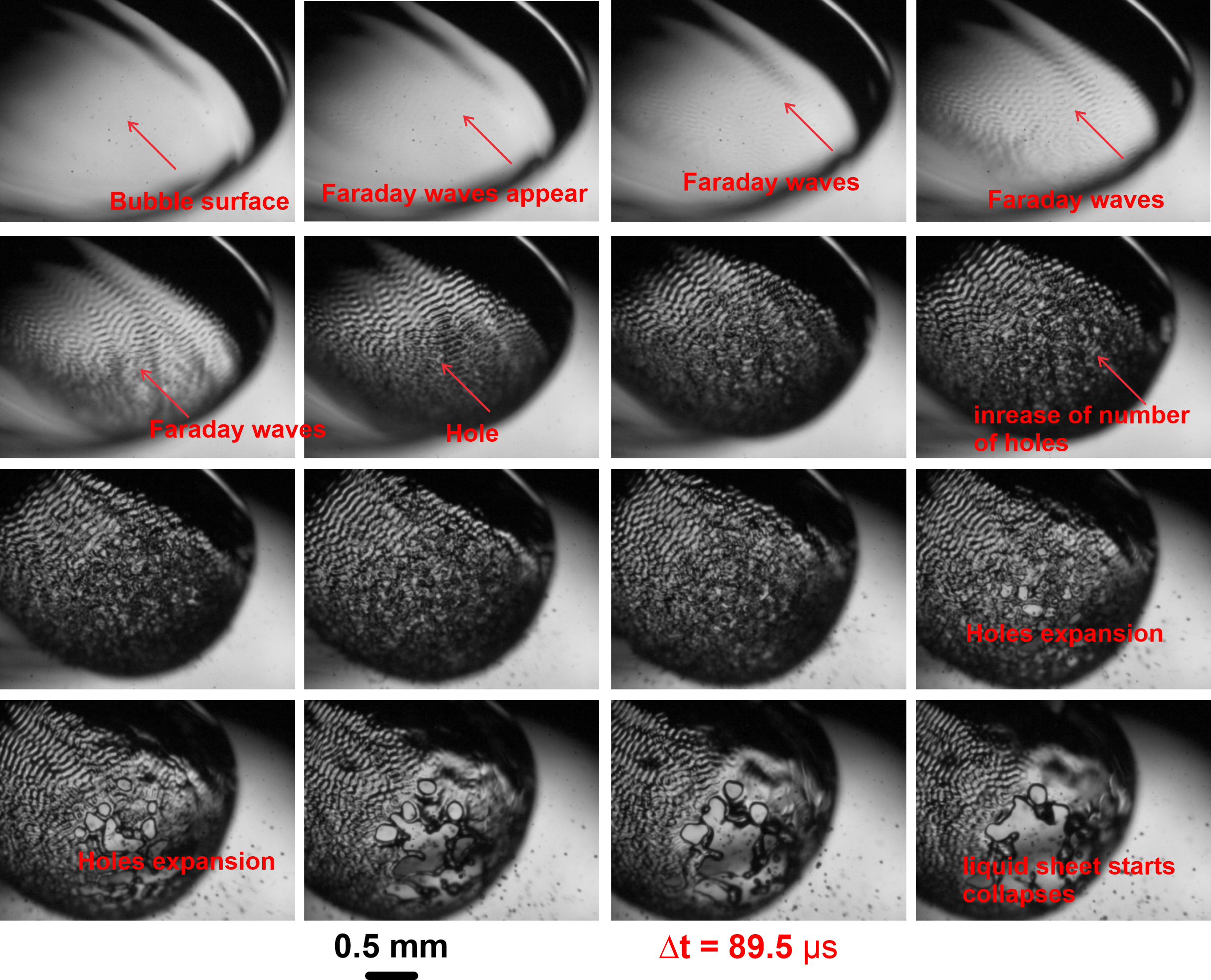}
\caption{\label{fig:bubbleburst} Instantaneous images of the growth of Faraday instability on the surface of the bubble and how that instability eventually resulted in the creation of a hole. The corresponding video is available at {\color{blue}\href{https://drive.google.com/file/d/1sxzraTZwSqpOOLYveg4fDRNHGDFcdg48/view?usp=share_link}{video: Bubble instability}}}
\end{figure*}

The observation of the bubble breakup is reported in section \ref{Bubble breakup}. The bubble breakup occurs for 1.56 $\le$ We $\le$ 1.92. The flattened droplet undergoes thinning followed by atomization at the edge. The liquid sheet starts to bend downwards leading to bag formation or bubble formation depending upon the initial size of the droplet (We). When the droplet size is increased such that during bag formation, its thickness does not reduce to an extent to admit sub-harmonic conditions, the liquid sheet takes a near bubble (see Figure \ref{fig:bubble}) like structure. The bubble further expands due to the pressure differences between the inside and outside of the bubble resulting in a further decrease in its thickness. As the thickness reduces to attain harmonic conditions, non-linear Faraday waves appear on the surface of the liquid as shown in Figure \ref{fig:bubbleburst}. These non-linear waves which are quite different from bag and umbrella breakup can be observed by focusing on the surface of the bubble. The measured wavelength is equal to 65 $\mu$m confirming it is a Faraday wave. This results in hole formation due to ejection of droplets due to Faraday instability, ligament formation due to hole rearrangement, and the breakup of ligaments due to RP instability as discussed in the section \ref{mechanismbag}. The rim can then dramatically close back and the bubble can show one more round of bubble expansion, finally leading to complete atomization by a combination of Faraday and RP instabilities.

The multistage breakup takes place for larger droplets with We $>$ 1.92. In this case, bag formation commences during equatorial breakup. The duration of an equatorial breakup is very small for a multistage breakup. Thus a very small volume of the droplet is lost due to the ejection of the liquid at the edge. The rim thickness increases as the liquid flowing from inside the droplet accumulates at the edge. As a consequence, instead of closing like a bubble, the bag's rim begins to expand in a radial direction. The breakup occurs in stages similar to the equatorial breakup. The Faraday wavelength is equal to 65 $\mu$m at the thinnest region indicating harmonic condition. The rim deformation in the azimuthal direction and its wavelength is equal to 0.34 mm. This rim deformation is triggered through RT instability, which is similar to equatorial breakup. The complete explanation is shown in the given video:{\color{blue}{\href{https://drive.google.com/file/d/1h_FLeFUPxIOZH5tiadkHMaC5fECerrI_/view?usp=share_link} {multistage}}}. This video is presented after bag formation.

\section{Conclusions}
An experimental investigation of the breakup dynamics of a single droplet levitated in the acoustic field is carried out for different sizes of the droplet (thereby changing the We). Both top and side view imaging are performed at a significantly higher frame rate (4000 fps to 140k fps) to understand the mechanism underlying the breakup process and visualize the associated interfacial instabilities. 
The droplet deforms with an increase in We without any atomization for We $\le$ 1.36. For  We $>$ 1.36, the droplet first transforms into a cylindrical liquid sheet during the stretching regime followed by the thinning at the edge. During the stretching period, a slow increase in velocity is observed, while the thinning regime exhibits an abrupt increment. Our theoretical model agrees well with the experimental results for the stretching regime. The radial expansion in thinning regime is highly non-linear and an empirical relationship based on $W_{eL}$, We, and nondimensional time scale (T) that explains both the stretching as well as thinning regimes is proposed. The thickness reduces at a faster rate in the thinning regime. As the thickness reduces to the order of the harmonic condition, the sheet vibrates in the vertical direction. A non-linear (Faraday) wave appears on the surface of the thin liquid sheet. The Faraday instability causes the ejection of smaller droplets perpendicular to the sheet. This droplet ejection creates the hole resulting in perforation of the liquid sheet. Then, the liquid rim is detached from the liquid sheet and breaks due to RP instability. The capillary (sinuous) wave generates due to the vertical vibration of the liquid sheet. The vertical vibration induces radial acceleration and deceleration towards the liquid-air interface (rim) resulting in the formation of the radial ligaments because of the development of RT instability. The sizes of the droplets, generated during the process, are related to the different types of interfacial instabilities that develop (Faraday, RT and RP instability) during equatorial breakup. The results of the experimental measurement show that although the pattern of the Faraday waves changes throughout all regimes, their wavelength remains constant. This suggests a parametric instability of the imposed wavelength on the sheet, $\lambda\sim U/f$. The bag formation commences during the atomization at the equatorial region. The liquid sheet diameter is larger than the levitation capacity of the acoustic levitator. Therefore, the liquid sheet bends downwards at the edges due to gravity. The radial pressure distribution along the sheet diameter causes the bag formation. The secondary breakup of the liquid sheet depends on the different types of the bag formed based on the We. The umbrella breakup for 1.36 $\le$ We $\le$ 1.45, bag breakup for 1.45 $<$ We $\le$ 1.56, bubble breakup for 1.53 $<$ We $\le$ 1.92  and multi-stage breakup for We $>$ 1.92 are observed experimentally. The mechanism of the breakup of the liquid sheet at different modes is discussed in detail. Experimentally measured wavelength is compared with calculated from the dispersion relation of the corresponding instability. \\

The work presents a myriad of complex phenomena that present significant challenges in theoretical predictions of the various involved instabilities, further exacerbated by the transient nature of the process. The work, however, demonstrates a variety of phenomena that could be harnessed judiciously while designing a large-scale atomizer based on acoustic fields.

\appendix
\section{}\label{appA}

\subsection{Basics of standing sound wave}\label{Dvelocity}
Consider an incident sound wave of frequency $f=40 kHz$, such that $\omega=2 \pi f$. The wavelength of sound is given by the distance the sound covers in 1 cycle i.e $1/f$ time. Thus $\lambda=2 \pi/k=c_o/f$, where $c_o$ is the velocity of sound, that is $c_o=\omega/k$. For the present case, $\lambda=$8.5 mm. The trap has an end-to-end distance $L=$ 90 mm, the number of the nodes where the droplet can be levitated (in the absence of gravity) can be obtained by $z_n = (n \lambda/2)+(\lambda/4)$. The number of the nodes (n) for the length L=$z_n$ is equal to $\sim$ 20.  This is exactly equal to our calculation from simulation using Comsol multiphysics software.

The actual conditions in the tiny-lev levitator are a little different due to multiple emitters and receptors. The incident pressure wave associated with a standing sound wave is given by (\cite{andrade2018review}),
\begin{equation}
    P_1= P_0 cos(kz) cos(\omega t)
\end{equation} 
\begin{equation}
    \rho_1= P_0/c_o^2 cos(kz) cos(\omega t)
\end{equation} 

such that $z_n=L= (n \lambda/2)+(\lambda/4)$
\begin{equation}
    k = \frac{ \pi }{2L} (2n+1)
\end{equation}. 
In a standing wave, the nodes occur at $z_n = (n \lambda/2)+(\lambda/4)$. For a single node, the node occurs at $z_n=L_1=3\lambda/4$.
The associated velocity with the standing sound wave that satisfies $\rho_o \frac{\partial u_1}{\partial t}=-\nabla P_1$, is given by, 
\begin{equation}
u_1=\frac{P_0}{\rho_0 v_0} sin(kz) sin(\omega t)
 = \frac{P_0}{\rho_0 v_0}  sin(\omega t))
\end{equation}
The RMS velocity associated with this pressure $<u_1^2>$
\begin{equation}
U^2 = \langle u_1^2\rangle =\bigg(\frac{P_0}{\rho_0 v_0}\bigg)^2     \frac{\int_{0}^{T=\frac{2\pi}{\omega}} sin^2(\omega t) \,dt  }{\frac{2\pi}{\omega}}
\end{equation}
\begin{equation}
 U =\frac{P_0}{\rho_0 v_0}  \frac{1}{\sqrt{2}}
\end{equation}

\subsection{Derivation of Net Radiation Pressure}\label{Derivation of pressure variation}
 The Bernoulli equation associated with the sound wave second order gives,
 \begin{equation}
\nabla <P_2>=-<\rho_1 \frac{\partial v_1}{\partial t}>-\rho_0 <v_1.\nabla v_1>
\end{equation}
 The time average radiation pressure can be given as,
\begin{equation}
\langle P_2 \rangle =\frac{\langle P_1^2 \rangle}{2 \rho_0 v_0^2}-\frac{\rho_0 \langle u_1^2 \rangle}{2}
\end{equation}
We get 
\begin{equation}
<P_2>=\frac{P_o^2}{4 \rho_0 c_0^2}\left( \cos^2{k z}-\sin^2{kz}\right)
\end{equation}

The associated Gorkov potential is given by,
\begin{equation}
U_{Gorkov}=\frac{P_o^2}{4 \rho_0 c_0^2}\left( \frac{\cos^2{k z}}{3}-\frac{\sin^2{k z}}{2}\right)
\end{equation}
Integrating over a sphere, one gets,
\begin{equation}
\label{frad}
F_{rad} {\bf e_z}=\frac{5 \pi R^3 k P_o^2}{6 \rho_0 c_0^2} \sin{2 k z} \, {\bf e_{z}}
\end{equation}

The droplet cannot be levitated at the exact node of the standing wave when gravity is present; however, it can be levitated just below the node point depending on the gravity force.

The force of gravity on a levitated sphere can be expressed as 
\begin{equation}
\label{gravity}
    F_g = -\rho_L g\frac{4 \pi R^3}{3}\,{\bf e_{z}}
\end{equation}

The z value comes out to be -0.3 mm after equating the equation \ref{frad} and equation \ref{gravity}, indicating that the droplet is levitated 0.3 mm below the node point.
\begin{equation}
\end{equation}

\subsection{Pressure around the droplet in the presence of a droplet}
\begin{equation}
    U^2 = U_x^2 + U_y^2 +U_z^2
\end{equation}

  $U_x = -(\gamma/2)$ x, $U_y = -(\gamma/2)$ y, $U_z = -\gamma$ z, where, $\gamma= f \frac{U}{d_0}$

\begin{equation}
   P_a(0,z)\approx  P_a(0) = P_a(r,z) + \frac{\rho_0 U^2}{2}
\end{equation}

\begin{equation}
    P_a(0) = P_a(r,z) + \frac{\rho_a \gamma^2}{2}  (U_x^2 + U_y^2 +U_z^2)
\end{equation}

\begin{equation}
    P_a(0) = P_a(r,z) + \frac{\rho_a f^2 U^2}{2 d_0^2}  (z^2+\frac{r^2}{4})
\end{equation}
\begin{equation}
    P_a(r,z) = P_a(0) - f^2\frac{\rho_a U^2}{2 d_0^2}  z^2- f^2 \frac{\rho_a U^2}{8 d_0^2}  r^2
\end{equation}

\subsection{Sound pressure distribution  inside the acoustic levitator}
\label{sound pressure}

COMSOL Multiphysics (COMSOL Inc.) software is used to simulate the pressure field distribution in the acoustic field. The Pressure Acoustics, Frequency Domain model is selected.

The following governing equation has been solve to get the pressure acoustic field distribution. 

\begin{equation}
\nabla \cdot \biggl( -\frac{1}{\rho}( \nabla p_t-q_d) \biggl) -\frac{k_{eq}^2}{\rho} p_t = Q_m
\end{equation}

\begin{equation}
k_{eq}^2 = \biggl( \frac{\omega}{c} \biggl)^2 - k_z^2 
\end{equation}

Where $Q_m$ and q are mono pole and dipole sources. $\omega = 2\pi f$, $f$ = 40k Hz. The geometry for the computational domain has been chosen according to the experimental setup. The two spherical cup of diameter 4.5 cm have been used at the top and bottom as acoustic field sources. The center distance between to cup is 9 cm. 

\begin{figure*}
\centering
\includegraphics[width=1\textwidth]{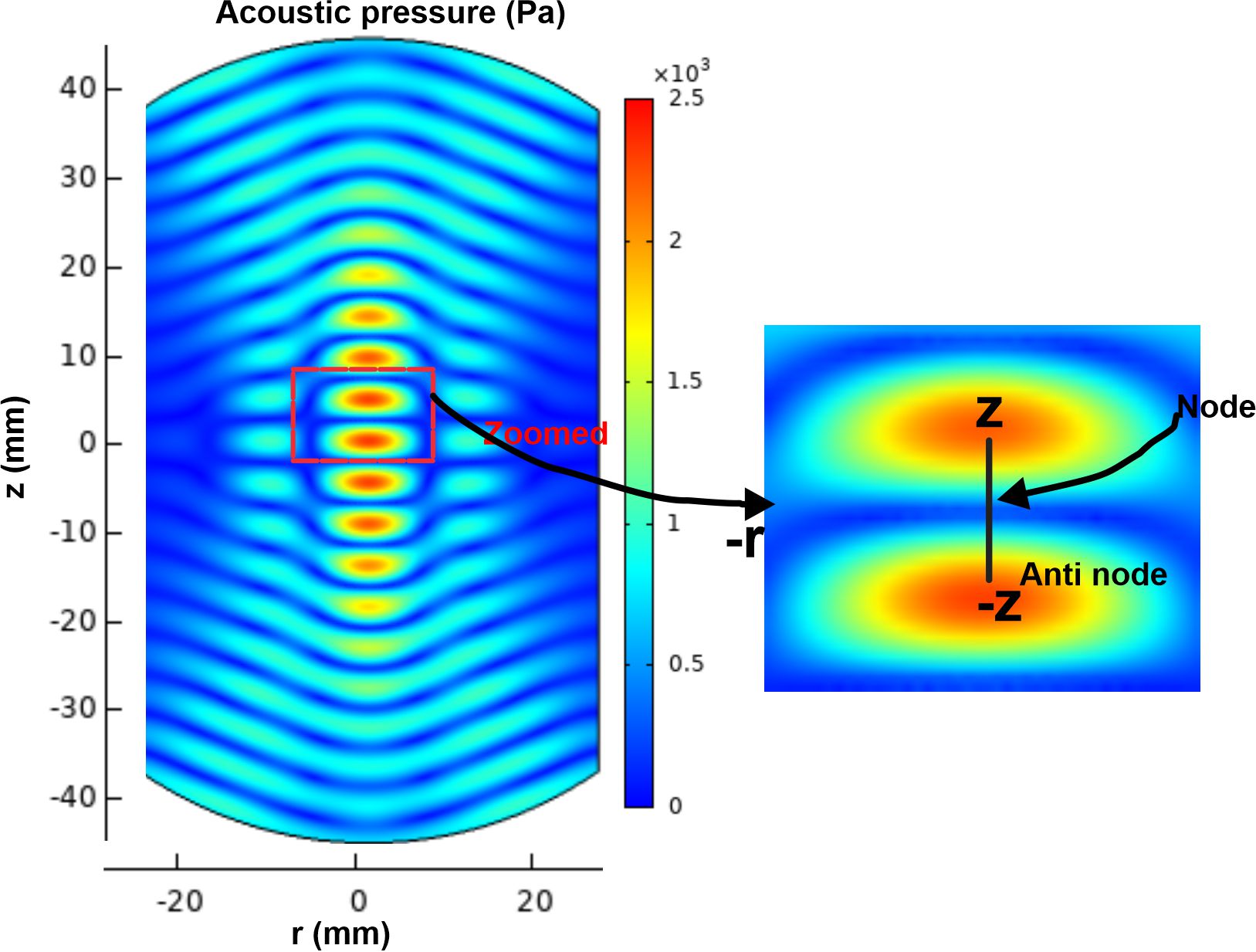}
\caption{\label{fig:soundpressure} Acoustic pressure field distribution inside the acoustic levitator.}
\end{figure*}
 Figure \ref{fig:soundpressure} shows the Acoustic pressure field distribution inside the acoustic levitator. The peak amplitude (pressure at anti node) is equal to 2400 Pa. The number of nodes are equal to 20. 

\subsection{Pressure re-distribution due to the presence of a particle in an acoustic field}
\label{acoustic thinning}
\begin{figure*}
\centering
\includegraphics[width=1\textwidth]{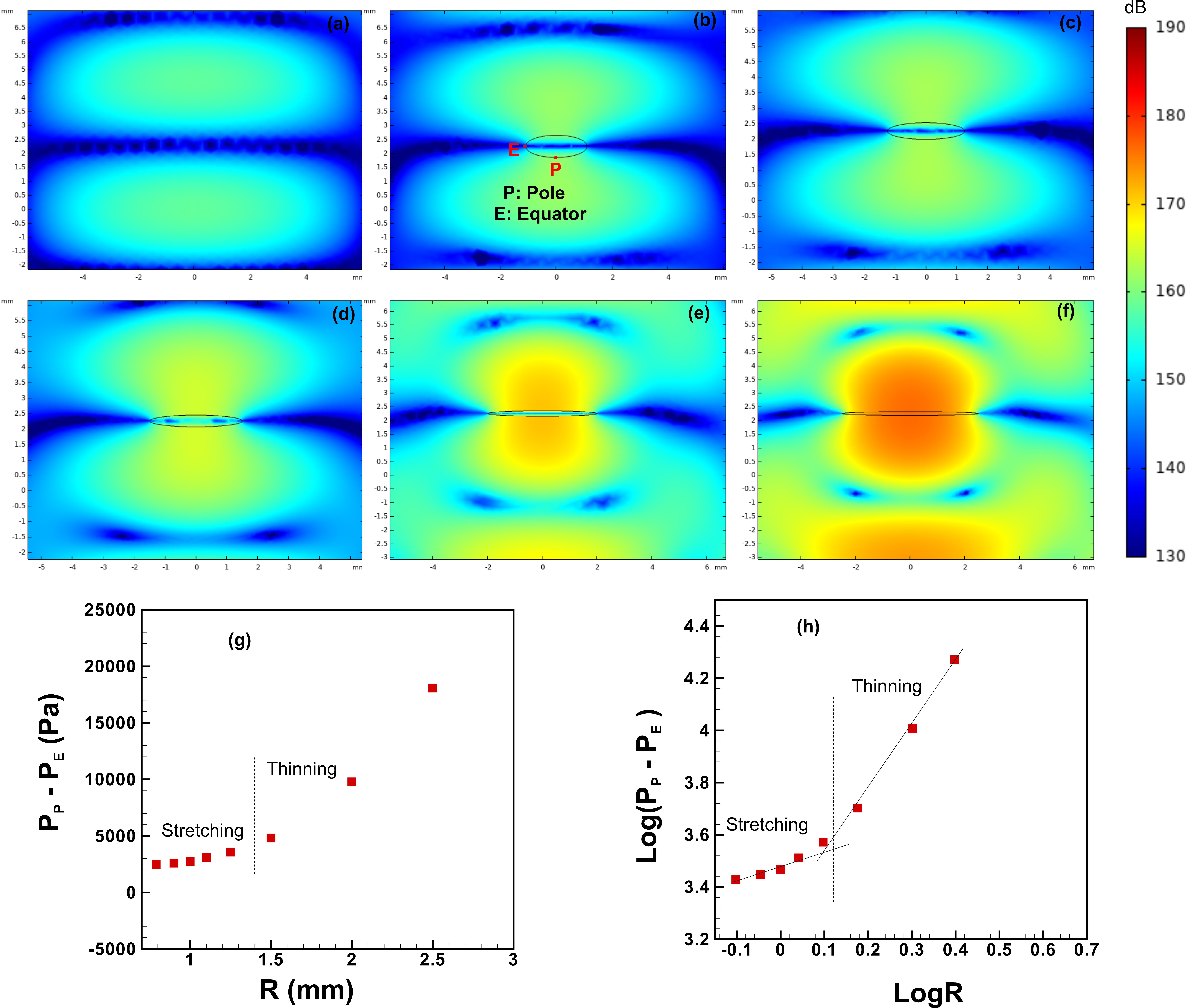}
\caption{\label{fig:comsol} Acoustic pressure field variation during deformation of the droplet levitated on the node of standing wave: (a) without drop (b) 2R = 2.2 mm and h = 0.8 mm, (c) 2R = 2.5 mm and h = 0.54 mm, (d) 2R = 3 mm and h = 0.38 mm, (e) 2R = 4 mm and h = 0.2 mm and (f) 2R = 5 mm and h = 0.14 mm, (g) $P_P - P_E$ vs R and Log($P_P - P_E$) vs Log R. Where $P_E$ and $P_P$ are the acoustic pressure at the equator and pole region, respectively. }
\end{figure*}

The presence of a particle (liquid droplet in this case), significantly and non-trivially alters the ambient undisturbed pressure distribution due to the standing sound wave. We have used the COMSOL multiphysics software to get the acoustic radiation pressure distribution around a droplet placed in a standing wave acoustic field. Our experimental observation suggests that the droplet shape changes with time in the standing wave. Therefore,  different aspect ratios of the deformed droplet were chosen according, to the experimental results, which cover the different regimes of dynamics of droplets in the acoustic field. The density, surface tension, and sound speed of the droplet are 830 $kg/m3$, 22.5 $mN/m$, and 1540 $m/s$, respectively. The obtained simulation results are presented in Figure \ref{fig:comsol}. Figure \ref{fig:comsol} (a) shows the distribution of the radiation acoustic pressure without any droplet. As the droplet is placed in a standing wave, the sound waves are scattered around the droplet such that the maximum pressure is at the pole region ($P_P$) and the minimum pressure is around the equatorial region ($P_E$) (Figure \ref{fig:comsol} (b)). The corresponding change in the pressure difference between the Pole and Equator ($P_P - P_E$)  for different aspect ratio is presented in Figure \ref{fig:comsol} (g), where P and E indicate the Pole and Equator, respectively, as marked in Figure \ref{fig:comsol} (b). The figure shows an increase in $P_P - P_E$ with an increase in $R$ and corresponding reduction in $h$, during the radial expansion of the droplet such that $P_P - P_E$ $\simeq$ $R^{1/2}$ during stretching regime and  $P_P - P_E$ $\simeq$ $R^{11/5}$ during thinning regime (see Figure \ref{fig:comsol} (f)) . \\

The nondimensional acoustic pressure on the surface of the droplet is given by \cite{shi1996deformation} in the form of acoustic velocity potential. The $U = -\nabla \phi$ and first order acoustic pressure $P = \rho_a \frac{\partial \phi}{\partial t}$. The velocity potential is nondimensionalized by the velocity potential of incident wave $\phi_0$. 

\begin{equation}
\label{eqnthinning}
    P_r = \frac{1}{K^2} (n.\nabla \phi)^2-\frac{1}{2} \phi^2
\end{equation}

Where, $K = \frac{k d_0}{2}$, k is the acoustic wave number and n is the unit normal vector on the surface of the droplet. 

As the droplet placed in the standing acoustic field, the acoustic wave scattered around the droplet and it experiences the non uniform radiation stress which causes the deformation of the droplet against the surface tension force. This is also reflected in our simulation results as shown in Figure \ref{fig:comsol} for without (a) and with (b) droplets in the standing acoustic field. The deformed shape further modifies the scattered acoustic field and radiation stress. This process continues until the transient energy is damped out whereafter it reaches a steady deformation. That indicates that in order to maintain equilibrium, the surface tension force is increased to the order of the radiation force by adjusting the curvature of the droplet.

The first term on the right-hand side of the equation \ref{eqnthinning} denotes the suction force in the outer normal direction and the second term denotes the squeezing force in the opposite direction. The suction force increased as the droplet becomes oblate shape under the effect of the acoustic field which causes the further flattening of the droplet. And at the same time, the squeezing force on the droplet is also increased (\cite{shi1996deformation}) which is similar to our simulation results. Once the droplet gets flattened, the radiation pressure becomes equal to the pressure inside the liquid sheet (see Figure \ref{fig:comsol} (e) and  Figure \ref{fig:comsol} (f)). The equilibrium can only be achieved if the surface tension, suction stress, and internal pressure balance each other. The surface tension force is proportional to the curvature ($\frac{1}{h}$) of the flattened droplet. The suction stress is proportional to the $\frac{P_0^2 R}{h}$ (\cite{lee1991static}). For the smaller value of the h, the internal pressure can be ignored because at this time the radiation pressure and pressure inside the liquid sheet become equal (see Figure \ref{fig:comsol} (d). A similar argument is also predicted by \cite{lee1991static}. Therefore, the dynamics at the rim are the competition between the surface tension and the suction stress. As the $P_0^2 R$ becomes larger during flattening, the surface tension force is no longer able to balance the suction force. The increased velocity at the equatorial region of the flattened droplet creates a higher suction. At this time, there is an insufficient increase of the squeezing force because of equal pressure. Therefore, the suction stress is completely dominated over the surface tension which pulls the liquid resulting in thinning at the equatorial region (\cite{lee1991static}). 

{\color{red}

\begin{table}
\caption{\label{table:f}. Stretching factor (f) for different We.}
\begin{center}

\begin{tabular}{ |c|c| } 

 \hline
 We &f \\ 
 \hline
 1.37 &9.2 \\ 
 1.45&9.28 \\ 
 1.56 &9.89\\
 1.92 & 9.93 \\ 
 2.4 &12\\ 
 \hline
\end{tabular}
\end{center}
\end{table}

}

\bibliographystyle{jfm}
\bibliography{jfm-instructions}

\end{document}